\documentclass[aps,pre,superscriptaddress,nofootinbib,twocolumn,10pt]{revtex4-2}
\usepackage{stmaryrd}
\usepackage{textcomp}
\usepackage{enumerate}
\usepackage{setspace}
\usepackage{amssymb}
\usepackage{amsthm}
\usepackage{amsmath}
\usepackage{graphicx}
\usepackage{extarrows}
\usepackage{marvosym}
\usepackage{empheq}
\usepackage{latexsym}
\usepackage{endnotes}
\usepackage{fontenc}
\usepackage{color}
\usepackage{comment}
\usepackage{url}
\usepackage{float}

\usepackage[colorlinks=true, allcolors=blue]{hyperref}

\newtheorem{theorem}{Theorem}

\newtheorem{lemma}[theorem]{Lemma}
\newtheorem{claim}[theorem]{Claim}
\newtheorem*{claim*}{Claim}
\newtheorem{corollary}[theorem]{Corollary}

\newtheorem{fact}[theorem]{Fact}

\newtheorem{proposition}[theorem]{Proposition}

\theoremstyle{definition}

\newtheorem*{definition*}{Definition}

\theoremstyle{remark}

\newtheorem{remark}[theorem]{Remark}
\newtheorem*{remark*}{Remark}

\begin{document}

	\newcommand\marginal[1]{\marginpar{\raggedright\parindent=0pt\tiny #1}}
	
	\theoremstyle{remark}

	\newcommand{\supp}{\mathrm{supp}}
	\def\eps{\varepsilon}
	\def\HH{\mathcal{H}}
	\def\E{\mathbb{E}}
    \def\D{\mathbb{D}}
	\def\C{\mathbb{C}}
	\def\R{\mathbb{R}}
	\def\Z{\mathbb{Z}}
	\def\N{\mathbb{N}}
	\def\PP{\mathbb{P}}
	\def\l{\lambda}
	\def\s{\sigma}
	\def\t{\theta}
	\def\a{\alpha}
	\def\la{\langle}
	\def\ra{\rangle}	
	\def\Xt{\widetilde{X}}
	\def\Pt{\widetilde{P}}
	\def\Var{\mathrm{Var}}
	\def\PV{\mathrm{PV}}
	\def\NV{\mathrm{NV}}
	\def\NN{\mathcal{N}}
	\def\CC{\mathcal{C}}
    \newcommand{\bX}{\mathbf{X}}
    \newcommand{\dH}{d_{\mathrm{H}}}
    \newcommand{\CL}{\mathbf{CL}}
    \newcommand{\dist}{\mathrm{dist}}
    
	\def\cA{\mathcal{A}}
	\def\cQ{\mathcal{Q}}	
	\def\cC{\mathcal{C}}
	\def\F{\mathcal{F}}
	\def\tm{\tilde{\mu}}
	\def\ts{\tilde{\sigma}}
	\def\L{\Lambda}
	\def\Q{\mathcal{Q}}
	\renewcommand{\P}{\mathbb{P}}
	\renewcommand{\Re}{\operatorname{Re}}
	\newcommand{\Kt}{\widetilde{K}}
	\newcommand{\mt}{\widetilde{\mu}}
	\newcommand{\Int}{\operatorname{int}}
	\newcommand{\Ext}{\operatorname{ext}}
	\newcommand{\nut}{\widetilde{\nu}}
	\newcommand{\mes}{\mathrm{mes}}
	\newcommand{\Cov}{\operatorname{Cov}}
	\newcommand{\cE}{\mathcal{E}}
	\newcommand{\open}{\mathcal{V}}
    \newcommand{\Vol}{\operatorname{Vol}}
    \newcommand{\Phic}{\Phi_{\mathrm{c}}}
    
    \newcommand{\marcus}[1]{{\color{red}{ [MM: #1]}}}
    \definecolor{dukeblue}{RGB}{1,33, 105}
    \newcommand{\pc}[1]{{\color{dukeblue}{\small\bf [PC: #1]}}}

	\title{On the critical value of void percolation in high dimensions}

\author{Patrick Charbonneau}
\affiliation{Department of Physics, Duke University, Durham, North Carolina 27708, USA}
\affiliation{Department of Chemistry, Duke University, Durham, North Carolina 27708, USA}

\author{Marcus Michelen}
\affiliation{Mathematics Department, Northwestern University, 2033 Sheridan Road, Evanston, Illinois 60208, USA}

\begin{abstract}
In the random Lorentz gas, a point tracer moves through spherical obstacles  of (reduced) density $\Phi$ centered on a Poisson point process in $d$-dimensional space, $\mathbb{R}^d$, and becomes localized once the vacant space stops percolating, at $\Phi_{\rm p}(d)$. The dimensional scaling of this void percolation threshold has remained elusive. Prior bounds, conjectural ones included, ranged from $\Omega(1)$ to $\mathcal{O}(d\ln d)$, leaving even the scaling undetermined. We here prove that $\Phi_{\rm p}(d)=\Theta(d)$, and in particular that $\tfrac12\,d \le \Phi_{\rm p}(d) \le 5\,d$ for all sufficiently large $d$. The upper bound counts hierarchical families of escape waypoints called tree embeddings; the lower bound is via a weighted second-moment estimate on an explicit ensemble of ballistic escape routes. A linear scaling of $\Phi_{\rm p}(d)$ excludes the $d\ln d$ growth suggested by analogy with other systems in the random first-order transition universality class and is consistent with the numerical estimate $\Phi_{\rm p}/d \to 3.40(5)$. Because the mean-field dynamical transition is likewise linear, $\Phi_{\rm d}\simeq 2.40\,d$, our result fixes the activated window $[\Phi_{\rm d},\Phi_{\rm p}]$ as a bounded—rather than diverging—correction to mean-field dynamical arrest in sufficiently high $d$.
\end{abstract}
	
	\maketitle

\section{Introduction}
 
The dynamical mean-field theory (DMFT) of simple glasses provides a
first-principles and quantitative account of the dynamical slowdown described qualitatively by the mode-coupling theory of liquids~\cite{parisi2019,manacorda2020,charbonneau2024}. The approach also brings systems as diverse as spin glasses, hard-sphere fluids, and the random Lorentz gas (RLG) under the common mean-field umbrella of the random first-order transition (RFOT) scenario~\cite{Biroli2021,biroli2021b,manacorda2022}. On its own, however, DMFT offers but limited insight into the physics of activated processes, which are key to understanding the finite-dimensional behavior of these various systems~\cite{berthier2010}. Although methodological advances have steadily sharpened this challenging problem over the last decade~\cite{baity-jesi2018,ros2019,ros2020,rizzo2021,ros2021,pacco2025,charbonneau2026}, conceptual clarity and controlled benchmarks both remain in short supply.

\begin{figure*}
    \centering
    \includegraphics[width=\linewidth]{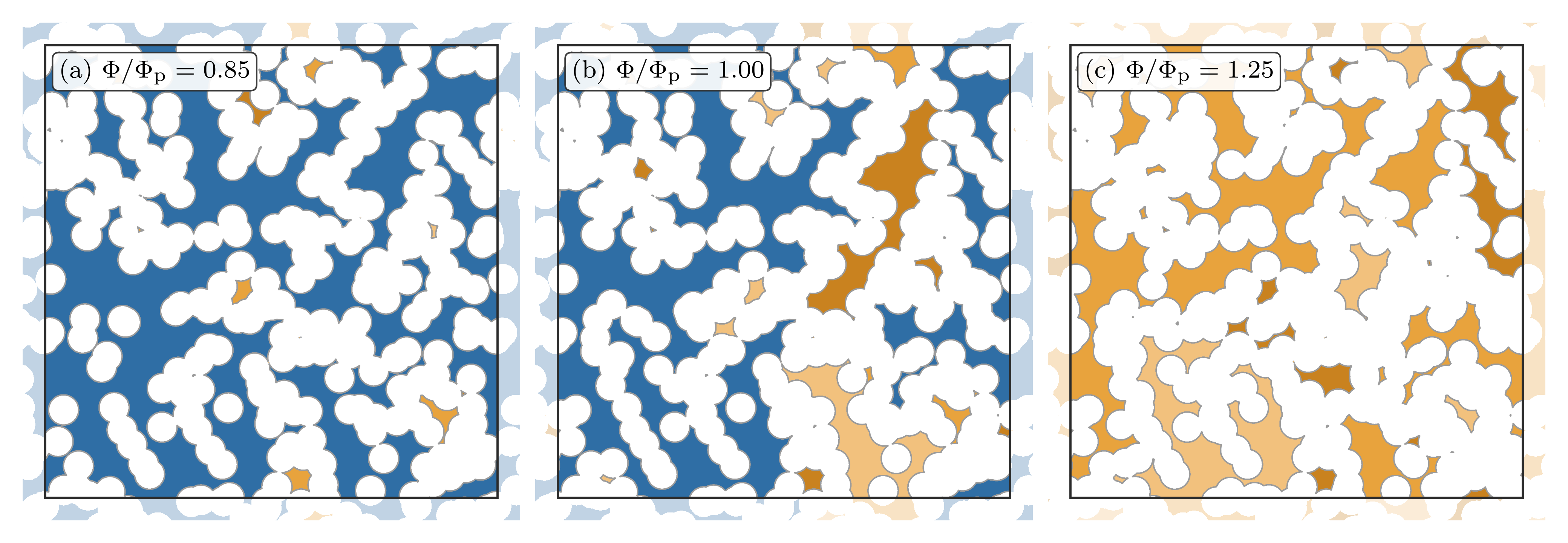}
    \caption{Sample $d=2$ configurations of obstacles of radius $\sigma$ (white) centered on the points of a Poisson point process of reduced density $\Phi$ (see Eq.~\eqref{eq:Phidef}), within a periodic box of side $L = 30\sigma$ (with faint periodic images surrounding each box). The vacant space available to a RLG tracer is decomposed into the component that wraps the torus (blue) and the finite cavities (orange, tinted individually). (a) Below threshold, $\Phi = 0.85 \Phi_\mathrm{p}$, the void forms a system-spanning network in which a tracer could diffuse without bound. (b) At threshold, $\Phi_\mathrm{p} = 1.12808737(6)$, the wrapping component is marginal. (c) Above threshold, $\Phi = 1.25 \Phi_\mathrm{p}$, no component wraps: the vacant space has fragmented into disconnected pockets and every tracer is localized. Note that in $d = 2$ the vacant and obstacle phases are dual, so $\Phi_\mathrm{p}=\Phi_\mathrm{c}$, a degeneracy lifted for $d\geq 3$.}
    \label{fig:setup}
\end{figure*}
 
In this context, the conceptual dichotomy presented by the RLG provides a particularly promising testbed.
On the one hand, according to the $d\to\infty$ DMFT, the RLG point tracer, which evolves among Poisson-distributed spherical obstacles, should become localized when the (reduced) density of obstacles crosses the dynamical transition, $\Phi_\mathrm{d} \simeq 2.40\,
d$~\cite{Biroli2021,biroli2021b}. On the other hand, the RLG has long been understood to be the continuum-space counterpart of the ant-in-the-labyrinth problem for lattice percolation, so the point tracer should become localized at the void percolation threshold, $\Phi_\mathrm{p}$ (see Fig.~\ref{fig:setup} and Sec.~\ref{sec:history}). A naive proposal
would have the two transitions coincide as $d\to\infty$, but that scenario quickly falters under the realization that the two localizations are of different physical natures: the percolation transition is continuous, whereas the dynamical transition is discontinuous~\cite{Biroli2021}. Because numerical computations suggest $\Phi_\mathrm{p}>\Phi_\mathrm{d}$ for sufficiently large $d$, it has instead been proposed that the approach to the void percolation threshold is associated with activated escape paths that lace together imperfect glass-like cages~\cite{jin2015,Biroli2021}. The escape probability would then acquire an instantonic large-deviation form in $d$, $P_\mathrm{escape}(t) \sim g(t e^{-d f(\Phi/d)})$. If this scenario is correct, then a study of RLG instantons could illuminate the nature of the void percolation transition as well as that of activated processes in RFOT systems more broadly.
 
A major obstacle to even articulating this program is that void percolation is relatively poorly understood in high $d$. As physically expected, mean-field theory fixes the critical exponents above the upper critical dimension for percolation,  $d_u=6$~\cite{Charbonneau2021}. The approach, however, says nothing about the location of the void percolation threshold as $d\to\infty$. And despite the purely geometrical nature of the problem, the underlying stochastic topology has long resisted rigorous analysis. Existing bounds---even those conjectured---are so loose that the plausibility of the instantonic scenario rests more on numerical inference from remarkably challenging computations~\cite{Charbonneau2021} than on a clear physical picture. The stakes are sharper than a mere tightening of constants. Because
$\Phi_\mathrm{d} \simeq 2.40\, d$, whether $\Phi_\mathrm{p}$ scales as
a constant $c$, as $d$, or as $d\ln d$ decides whether the activated window exists, closes to a fixed ratio, or widens without bound as $d\to\infty$.
 
In this article, we prove that $\Phi_{\mathrm{p}}(d) = \Theta(d)$ using mathematical tools developed from random sphere packings and coverings as well as high-dimensional percolation.  In Appendix \ref{sec:appendix} we further prove that $\Phi_{\mathrm{p}}(d) /d \in [1/2,5]$ for all $d$ sufficiently large. By contrast, numerics from prior works---detailed in Section \ref{sec:history} below---were not able to rule out the possibility that $\Phi_{\mathrm{p}}(d) = \Theta(d \log d)$. 
We therefore find the linear scaling in $d$ to be generally consistent with the instantonic scenario sketched above.  If the proposed activated regime exists, our upper bound confines it to densities of order $d$, below the $d \log d$ scale suggested by analogies with other RFOT systems.  
 
The rest of this article is structured as follows. Section~\ref{sec:history} reviews prior results on the connection between the RLG and void percolation and summarizes earlier efforts at bounding the latter. Lower and upper bounds are then obtained in Secs.~\ref{sec:lower} and~\ref{sec:upper}, respectively. A brief conclusion expanding on the physical implications of these results follows in Sec.~\ref{sec:conclusion}.

\section{Physical and mathematical background}
\label{sec:history}
 
In 1905, Lorentz introduced a minimal model of electronic conduction in solids:
a single point particle moving ballistically through---and undergoing elastic
reflections with---an array of fixed scatterers~\cite{Lorentz1905}. Lorentz,
however, prescribed no spatial arrangement for these obstacles. A few years later,
Paul and Tatyana Ehrenfest~\cite{Ehrenfest1911} fixed the obstacle geometry such that the combinatorics of collision sequences could be carried out explicitly: the wind-tree
form with immobile square obstacles, explicitly irregularly distributed over the plane with their diagonals parallel to the coordinate axes, and point particles restricted to the four velocity directions parallel to those
diagonals.\footnote{The ``wind-tree'' descriptor, by which the model is now known, does not appear in the original work; Hauge and Cohen~\cite{HaugeCohen1969} attribute it to Paul Ehrenfest's lectures instead.} 
Both models---the former specifically in its \emph{random}, or RLG, form---later became standard proving grounds for questions about the kinetic theory of gases, such as the validity of the Boltzmann equation~\cite{Gallavotti1969,Spohn1978} and the existence of a density expansion for the transport coefficients~\cite{brush1972kinetic,resibois1977}.
 
The latter question led, indirectly, to considerations of the dense regime, where void percolation is encountered. Because a naive density expansion of the inverse diffusion coefficient diverges term by term beyond the Boltzmann order, the divergence must be resummed~\cite{vanleeuwen1967,weijland1968}. To make this concrete, Hauge and Cohen in the late 1960s took up the two-dimensional wind-tree model with all tree configurations being equally probable, so that the squares may overlap freely---a Poisson, and hence a Boolean, model~\cite{HaugeCohen1967,HaugeCohen1969}. It is in the concluding remarks of that paper that the connection central to the present work is first articulated. Noting that with overlapping trees ``there is always a nonvanishing probability that a moving particle started in a random position between the trees is trapped in a finite volume,'' so that ``the displacement can never exceed finite bounds and nothing resembling a diffusive process is ever realized,'' Hauge and Cohen conjectured, ``in analogy with the percolation problem,'' that this probability increases with density and becomes unity above a certain critical density, ``which then denotes the end of the diffusive regime''~\cite[p.~411, n.~27]{HaugeCohen1969}.
 
A decade later, Alder and Alley~\cite{AlderAlley1978,Alley1979}, studying the $d=2$ RLG by molecular dynamics, simply claimed that ``the lowest density at which the diffusion coefficient vanishes is called the percolation point.'' That such a claim could be made without either physical or mathematical argument is intelligible in the light of what the percolation literature of the preceding decade had put in circulation. Ambegaokar, Halperin and Langer~\cite{Ambegaokar1971} had reduced hopping conduction to a percolation threshold by critical-path analysis; Kirkpatrick's numerical study of dilute resistor networks~\cite{Kirkpatrick1971,Kirkpatrick1973} had established a power law for lattice conduction near threshold, and argued that ``this behavior should be a general feature of transport near a percolation threshold in continuous media as well''~\cite[p.~574]{Kirkpatrick1973}; and de Gennes~\cite{deGennes1976} had recast percolative conduction as the diffusion of a tracer---an ant in a labyrinth---confined to a finite cluster below threshold. In lattice-based setups, this identification has since been placed on rigorous footing in precisely the high-dimensional limit of interest here: the Alexander--Orbach conjecture holds for percolation in high dimensions~\cite{KozmaNachmias2009}, and scaling limits have been established for the ant itself~\cite{BenArousCabezasFribergh2019}.

An explicit geometrical connection between percolation and dynamics for the RLG was supplied by Kertész a few years later~\cite{Kertesz1981}. By computing the threshold for void percolation between overlapping spheres directly,
the work clearly identified it as the density at which RLG localization sets in.
Kerstein~\cite{Kerstein1983} then proved the equivalence of the void percolation
problem for overlapping spheres with a network problem on the Voronoi tessellation
of the sphere centers---the mapping on which the numerical results quoted below
still rest. That the localization transition is quantitatively the void percolation transition was assessed numerically only much later, in $d=3$~\cite{Hofling2006} and $d=2$~\cite{Bauer2010}; the threshold and its criticality were determined up to $d=9$ a decade on~\cite{Charbonneau2021}.
 
From a stochastic topology viewpoint, the object in question is the vacant
component of a Poisson Boolean model: balls of radius $\sigma$ centered on the
points of a Poisson point process of intensity $\lambda$ in $\mathbb{R}^{d}$, with the complement of their union as the medium. Following Meester and Roy~\cite{MeesterRoy1996}, we consider the reduced density
\begin{equation}
  \Phi = \lambda\, \kappa_{d}\,\sigma^{d},
  \qquad
  \kappa_{d} = \frac{\pi^{d/2}}{\Gamma(1 + d/2)},
  \label{eq:Phidef}
\end{equation}
with $\kappa_d$ the volume of a $d$-dimensional unit ball.  The associated void volume fraction is then $e^{-\Phi}$. Write $\Phi_\mathrm{c}$ and $\Phi_\mathrm{p}$ for the thresholds of the occupied and of the vacant phase, respectively. The two quantities scale very differently as $d \to \infty$. For the occupied phase the asymptotics are settled: $\Phi_\mathrm{c} \sim 2^{-d}$ was proven by Penrose \cite{Penrose1996} in 1996; later, the asymptotic $\Phi_\mathrm{c} \sim 2^{-d}$ was argued physically by Torquato~\cite{Torquato2012} on the basis of his numerical results with Jiao for $d =2\ldots 11$~\cite{TorquatoJiao2012} and reproven by Anantharam and Baccelli~\cite{Anantharam2016}. For the void phase the corresponding question was, until recently, open in both directions: by comparison with dimension $2$, Penrose \cite{Penrose1996} proves $\Phi_{\mathrm p}(d) = \Omega(d^{-1})$ and no quantitative high-dimensional asymptotic upper bound was established; finite upper bounds in each fixed $d \geq 2$ are classical \cite[Chapter~4]{MeesterRoy1996}. The present state of the art was set in Ref.~\cite{Charbonneau2021}.
\begin{itemize}
\item\emph{Lower bound.} Section III C of that work argues that
$\Phi_\mathrm{p}(d) = \Omega(1)$ as $d \to \infty$. The argument is elementary in structure: the $d$-dimensional
void percolates if its restriction to a fixed $n$-dimensional hyperplane does, and
on such a hyperplane the obstacles form a Boolean model with random radii whose
limiting distribution has finite moments, for which the vacancy transition is
known to be non-trivial~\cite{Penrose2018,Gouere2008}.
\item\emph{Upper bound.} Section III B of that work predicts that
$\Phi_\mathrm{p}(d) \le \mathcal{O}(d \ln d)$. The construction bounds the probability
that a tracer can leave a spherical shell about its starting point through an
obstacle-free cap. The argument is conjectural in part because
it relies on the expected critical form $(1 - \Phi/\Phi_\mathrm{p})^{\beta}$ with
$\beta = 1$ for $d \ge d_{u} = 6$, which is standard percolation physics but is
not rigorously established for this model.
\item\emph{Numerics.} That work also finds its numerical results for $\Phi_\mathrm{p}$ in $d=2\ldots9$ to be reasonably consistent with both $\Phi_\mathrm{p}/d \to 3.40(5)$ and $\Phi_\mathrm{p}/(d\ln d) \to 0.98(5)$, the former giving a lower and the latter an upper estimate of the asymptotic growth. Barring major large $d$ surprises, one therefore expects $3.40\,d \;\lesssim\; \Phi_\mathrm{p} \;\lesssim\; d\ln d$. 
\end{itemize}
The wide gap between the proven $\Omega(1)$ and the conjectured $\mathcal{O}(d\ln d)$ is the precise sense in which the asymptotic scaling of $\Phi_\mathrm{p}$ remains open. Tightening it is the object of the rest of this work.  In particular, we here prove that $\Phi_{\mathrm p}(d) = \Theta(d)$.

\begin{theorem}
    There are universal constants $c,C > 0$ so that for all $d$ sufficiently large we have $$c d \leq \Phi_{\mathrm p}(d) \leq C d\,.$$
    \label{thm:bounds}
\end{theorem}

We note that ostensibly it could be the case that $\Phi_{\mathrm p}(d)/d$ fails to converge due to oscillations.  We conjecture that $\lim_{d \to \infty} \Phi_{\mathrm p}(d) / d$ exists, but have no method for proving it.  

\begin{remark}
    Our proof in the body of the paper is written in order to maximize readability and not to optimize the values of the constants $c$ and $C$; however, the proof can be altered with further numerical bounds to show that for sufficiently large $d$ one in fact has \begin{equation}\label{eq:explicit-constants}
        \frac{1}{2} \cdot d \leq \Phi_{\mathrm p}(d) \leq 5 \cdot d\,.
    \end{equation}
    We indicate in Appendix \ref{sec:appendix} the argument along with the  programs that support this result~\cite{RDR}.
\end{remark}

\section{Definition and set up}
\label{sec:setup}
Given a density $\Phi > 0$ and a dimension $d$, let $\bX$ be the Poisson point process of intensity $\lambda$ in $\R^d$.  We will take the radius $\sigma$ of the balls centered at the points of $\bX$ to be equal to the radius $r_d$ of the ball of volume $1$.  In the notation of Eq.~\eqref{eq:Phidef}, we have $\sigma = r_d = \kappa_d^{-1/d}$ and so $\Phi = \lambda$.  
By Stirling's formula, we recall that $r_d = \Theta(\sqrt{d})$.  We define \begin{equation}
	\open_\Phi = \R^d \setminus \bigcup_{x \in \bX} B(x,r_d)
\end{equation}
to be the vacant set, where we will often write $\open = \open_\Phi$. Here and below we write $B(x,s)$ for the closed ball, and thus $\open$ is open.  Setting $$\theta_d(\Phi) = \P(0 \text{ lies in an unbounded component of }\open_\Phi)$$
the critical threshold for void percolation is given by \begin{equation*}
    \Phi_{\mathrm p}(d) = \sup\{\Phi \geq 0: \theta_d(\Phi) > 0\}\,.
\end{equation*}

\section{Upper bound}
The main goal of this section is to prove the upper bound in Theorem \ref{thm:bounds}:
\label{sec:upper}
\begin{proposition} \label{prop:upper-bound}
	There is a universal constant $M > 0$ so that for $d$ sufficiently large and $\Phi = Md$ we have $$\theta_d(\Phi) = 0\,.$$
\end{proposition}

Throughout, we write $r = r_d = \kappa_d^{-1/d}$  for the radius of the ball of volume $1$ in $\R^d$.  
Define \begin{equation}\label{eq:choice-of-parameters}
    t = 10^{-10}\,, \qquad \rho = \sqrt{r^2 - t^2}\,,
\end{equation}
and note first that \begin{equation}\label{eq:rho-vol}
q_d = \Vol(B(0,\rho)) = \left(1 - \frac{t^2}{r^2}\right)^{d/2} \geq 1/2\,,
\end{equation}
where the last bound is due to the fact that $r^2 \geq d / (2\pi e)$.  

Our first step---inspired by a work of Dumer on the covering number of Euclidean space \cite{dumer2007covering}---is that instead of asking if a given point $x \in \R^d$ is uncovered, we ask if the ball $B(x,t)$ fails to be entirely covered; more technically, it will be simpler to bound this quantity if we give ourselves some wiggle room, and in particular we will ask for the event that $B(x,t)$ is covered by balls whose centers lie in $B(x,\rho)$.  For an arbitrary point $x \in \R^d$ we define \begin{equation}
    F_x = \left\{B(x,t) \not\subseteq \bigcup_{y \in \bX \cap B(x,\rho)} B(y,r) \right\}\,,
\end{equation}
which is our upper bound for the event $x \in \open$. We now upper bound the probability $\P(F_x)$.   One can adapt the arguments of Dumer \cite{dumer2007covering} to show this bound, but we use Wendel's formula instead~\cite{wendel1962problem}.   
\begin{lemma}\label{lem:one-point-estimate}
    There are absolute constants $C_0,c_0 > 0$ so that for $\Phi \geq C_0 d$ we have $$\P(F_x) \leq e^{-c_0\Phi}\,. $$
\end{lemma}
\begin{proof}
    Assume without loss of generality that $x = 0$ and set $N = |\bX \cap B(0,\rho)|.$  Note that $N$ is a Poisson random variable with $\E N = \Phi q_d \geq \Phi/2$ by Eq.~\eqref{eq:rho-vol}.  On the event $N = n$, let $Y_1,\ldots,Y_n$ denote the points of $\bX \cap B(0,\rho)$.  Note that if $F_0$ occurs, then there is some point $z \in B(0,t)$ for which $|z - Y_j| > r$ for all $j$.  Since $\|z\|_2^2 + \|Y_i\|_2^2 \leq t^2 + \rho^2 = r^2$, this implies that $\langle z, Y_j\rangle < 0$ for all $j$.  In particular, this implies that the unit vectors $\widehat{Y}_j = Y_j / \|Y_j\|_2$ all lie in a hemisphere.  Letting $\mathcal{E}_n$ denote the event that the unit vectors $\{\widehat{Y}_j\}_{j = 1}^n$ all lie in a common hemisphere, Wendel's formula \cite{wendel1962problem} provides the exact computation$$\P(\mathcal{E}_n) = 2^{1 - n} \sum_{j = 0}^{d - 1} \binom{n-1}{j}\,.$$
    For $n \geq 3d$, say, one obtains the asymptotic  $$\P(\mathcal{E}_n) \leq 2^{1 - n} \cdot d \cdot \binom{n}{n/3} \leq e^{-n/20}$$
    for sufficiently large $d$.  Because $N$ is a Poisson variable we have  $\P(N \leq \E N / 2) \leq e^{- \E N / 10}$ and so we have \begin{equation*}
        \P(F_x) \leq \P(N \leq \E N /2) + \P(\mathcal{E}_N \,|\,N \geq \E N/2) \leq e^{-\Omega(\Phi)}
    \end{equation*}
    where in the last bound we used that $\E N \geq \Phi/2.$
\end{proof}

It will be useful to record the following corollary:
\begin{corollary}\label{cor:conditional-thin-lens}
    For a set $U \subset B(x,\rho)$, define $F_x(U) = \{ B(x,t) \not\subseteq \bigcup_{y \in \bX \cap U} B(y,r) \}$.  Then for $C_0,c_0$ in Lemma \ref{lem:one-point-estimate} we have $$\P(F_x(U)) \leq e^{\Phi \Vol(B(x,\rho) \setminus U)} e^{- c_0 \Phi}\,.$$ 
\end{corollary}
\begin{proof}
    This follows from noting that $\P(\bX \cap (B(x,\rho) \setminus U) = \emptyset) = e^{-\Phi \Vol(B(x,\rho) \setminus U)}$.  Noting that $F_x(U) \cap \{\bX \cap (B(x,\rho) \setminus U) = \emptyset\} \subset F_x$ completes the proof after using Lemma \ref{lem:one-point-estimate}.
\end{proof}

Define \begin{equation}
    V_x = \{B(x,t) \cap \open \neq \emptyset\}
\end{equation}
to be the event that $B(x,t)$ contains an open point.  We want to estimate the probability that $V_x$ occurs for many well-separated $x$ simultaneously.  

\begin{figure*}
    \centering
    \includegraphics[width=\linewidth]{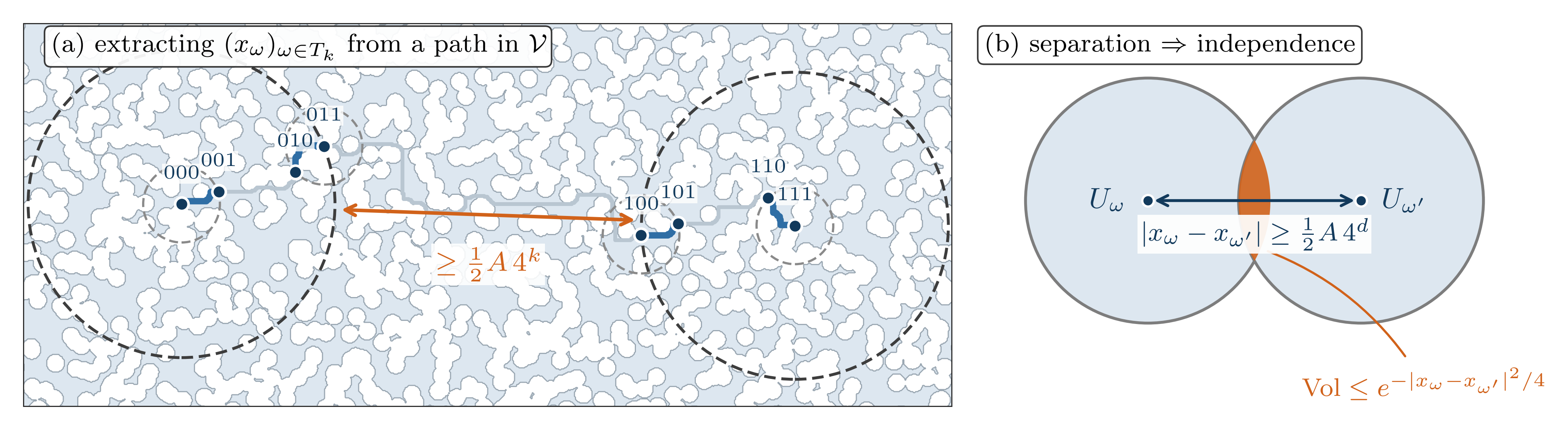}
    \caption{Construction of a tree embedding in $d=2$. (a) Lemma~\ref{lem:extract-tree} applied to a path $\eta$ in the vacant set $\open$ (obstacles white, $\open$ shaded), to scale for $k=3$ with $|\eta(0)-\eta(1)| = A\cdot 4^{k}$.  Dashed circles of radius $A \cdot 4^{k-1}$ about the endpoints fix the first exit and last entry; the retained pieces are dark, the discarded middle light.  Their radii being a quarter of the center separation, the retained pieces lie at least $\tfrac{1}{2}A \cdot 4^{k}$ apart (arrow), and each has endpoint separation $A \cdot 4^{k-1}$, so the construction is recursive with radii shrinking by $4$ per level---the origin of $4^{d(\omega,\omega')}$ in Eq.~\eqref{eq:tree-separation}. (b) Fact~\ref{fact:overlap-gaussian-decay} gives that the lens excised in forming $U_\omega$ has volume at most $e^{-|x_\omega-x_{\omega'}|^{2}/4}$ , so the separation of (a) leaves the $U_\omega$ pairwise disjoint and the $2^{k}$ events independent.}
    \label{fig:tree}
\end{figure*}

Ultimately, given a path in $\open$ we will find points along the path whose distances are approximately those of a binary tree.  In particular, define $T_k = \{0,1\}^k$ to be the leaves of the complete binary tree of depth $k$.  For two points $\omega, \omega' \in T_k$ define $d(\omega,\omega')$ to be so that the longest common prefix has length $k - d(\omega,\omega')$.  In particular, the distance in the tree from $\omega$ to $\omega'$ is exactly $2d(\omega,\omega')$ and each is exactly distance $d(\omega,\omega')$ from their most recent common ancestor.  For an absolute constant $A > 0$ to be chosen later, we call a tuple $(x_\omega)_{\omega \in T_k}$ a \emph{tree embedding} if (see Fig.~\ref{fig:tree}) \begin{equation}\label{eq:tree-separation} 
    |x_{\omega} - x_{\omega'}| \geq \frac{1}{4}A \cdot4^{d(\omega,\omega')}\,. 
\end{equation}

We note that similar tree embeddings appear in the literature on percolation, for instance in the work  on the vacant set of random interlacements~\cite{rath2015short}.  We specifically point the reader to \cite[Section~3]{rath2015short} for the relevant extraction, counting, and separation statements.

Throughout, we will use a basic bound on the decay of the overlap between balls in $\R^d$ (see e.g.\ \cite[Fact 2.3]{campos2023new}).

\begin{fact}\label{fact:overlap-gaussian-decay}
    For all $d$ and  $x,y \in \R^d$, we have $$\Vol(B(x,r) \cap B(y,r)) \leq e^{-|x - y|^2 / 4}\,.$$
\end{fact}

Together with Corollary \ref{cor:conditional-thin-lens}, Fact \ref{fact:overlap-gaussian-decay} will allow us to obtain approximate independence of the events $V_x$ across the points in a tree embedding.

\begin{lemma}\label{lem:tree-esimate}
    There are absolute constants $A,C_1,c_1 > 0$ so that the following holds for sufficiently large $d$.  If $\{x_\omega\}_{\omega \in T_k}$ is a tree embedding and $\Phi \geq C_1 d$ then \begin{equation*}
        \P\left(\bigcap_{\omega \in T_k} V_{x_\omega} \right) \leq \exp\left( - c_1 \Phi 2^k\right)\,.
    \end{equation*}
\end{lemma}
\begin{proof}
    For $\omega \in T_k$, define \begin{equation*}
        U_{\omega}  = B(x_\omega,\rho) \setminus \bigcup_{\omega' \neq \omega} B(x_{\omega'},\rho)\,,
    \end{equation*}
    which we note are pairwise disjoint.  Setting ourselves up to apply Corollary \ref{cor:conditional-thin-lens}, bound 
    \begin{align*}
        \Vol(B(x_\omega,\rho) \setminus U_\omega) &\leq \sum_{\omega' \neq \omega} \Vol( B(x_\omega,\rho) \cap B(x_{\omega'},\rho))\\
        &\leq \sum_{\omega' \neq \omega} \exp\left(- |x_\omega - x_{\omega'}|^2 / 4 \right) \\
        &\leq \sum_{j = 1}^k 2^{j} \exp\left( -A^2 16^{j} / 64 \right)\,,
    \end{align*}
    where in the second bound we used Fact \ref{fact:overlap-gaussian-decay} and in the last we used the definition of a tree embedding in Eq.~\eqref{eq:tree-separation}. We may thus choose $A$ large enough so that we have $\Vol(B(x_\omega,\rho) \setminus U_\omega) \leq c_0/2$ where $c_0$ is from Lemma \ref{lem:one-point-estimate}.  Applying Corollary \ref{cor:conditional-thin-lens} then shows that 
    \begin{equation}\label{eq:F-lens-UB}
        \P(F_{x_\omega}(U_\omega)) \leq e^{c_0 \Phi /2} \cdot e^{-c_0\Phi} = e^{-c_0 \Phi /2}\,.
    \end{equation}
    Because the sets $\{U_{\omega}\}_{\omega \in T_k}$ are mutually disjoint, the events $F_{x_\omega}(U_\omega)$ are mutually independent and satisfy $V_{x_\omega} \subset F_{x_\omega}(U_\omega).$  Multiplying Eq.~\eqref{eq:F-lens-UB} across all $2^k$ elements of $T_k$ completes the proof for $c_1 = c_0/2$ and $C_1 = C_0$.
\end{proof}

\subsection{Extracting a tree embedding and a net for tree embeddings}

We first show how to extract a tree embedding from a continuous curve by iteratively cutting it; see Figure \ref{fig:tree}.  We write $0^k$ and $1^k$ for the binary strings of length $k$ consisting only of $0$ and $1$ respectively.

\begin{lemma}\label{lem:extract-tree}
    Let $\eta:[0,1] \to \R^d$ be a continuous curve so that $|\eta(0) - \eta(1)|  = A \cdot 4^k.$ Then there are times $t_{\omega} \in [0,1]$ for $\omega \in T_k$ so that the points $x_\omega = \eta(t_\omega)$ satisfy $$\frac{1}{2} A \cdot 4^{d(\omega,\omega')} \leq |x_\omega - x_{\omega'}| \leq \frac{3}{2} A \cdot 4^{d(\omega,\omega')}$$
    with $t_{0^k} = 0$ and $t_{1^k} = 1.$
\end{lemma}
\begin{proof}
    The proof is via induction on $k$ where we note the $k = 1$ case follows from taking $t_0 = 0$ and $t_1 = 1$.   For the inductive step, define $\tau_0 = \min\{ t \in [0,1] : |\eta(t) - \eta(0)| = A \cdot 4^{k-1}\}$ and $\tau_1 = \max\{ t \in [0,1] : |\eta(t) - \eta(1)| = A \cdot 4^{k-1}\}$.  Define the curves $\eta_0,\eta_1:[0,1] \to \R^d$ via \begin{equation*}
        \eta_0(t) = \eta(t \tau_0), \qquad \eta_1(t) = \eta(\tau_1 + t(1 - \tau_1))\,,
    \end{equation*}
    and note that $\eta_0([0,1]) \subset B(\eta(0), A \cdot 4^{k-1})$ and $\eta_1([0,1]) \subset B(\eta(1), A \cdot 4^{k-1})$.  By the triangle inequality, we have that all points on the curve $\eta_0([0,1])$ are at least distance $2 \cdot A \cdot 4^{k-1} = \frac{1}{2} A 4^{k}$ from all points on $\eta_1([0,1])$; the upper bound is analogous.  Further, note that $|\eta_0(0) - \eta_0(1)| = |\eta_1(0) - \eta_1(1)| = A \cdot 4^{k-1}$.  Applying the inductive hypothesis to each of the curves $\eta_0$ and $\eta_1$ completes the proof.
\end{proof}

We now will construct a net for tree embeddings.  We start with a basic net construction on $\R^d$:

\begin{fact}\label{fact:basic-net}
    For each $\delta \in (0,1/4)$, there is a $\delta$ net $\mathcal{N}$ of $\R^d$ with $0 \in \mathcal{N}$ so that for all $x \in \R^d$ and $R > 0$ we have \begin{equation*}
        |\mathcal{N} \cap B(x,R)| \leq (1 + 2 R/\delta)^d\,.
    \end{equation*}
\end{fact}
\begin{proof}
    Let $\mathcal{N}$ be any saturated sphere packing of balls of radius $\delta/2$ that contains $0$.  Then $\mathcal{N}$ is a $\delta$-net since it is a saturated packing.  Further, since $\{B(y,\delta/2)\}_{y  \in \mathcal{N} \cap B(x,R)}$ are disjoint with $\bigcup_{y  \in \mathcal{N} \cap B(x,R)} B(y,\delta/2) \subset B(x, R + \delta/2)$ we have \begin{equation*}
        |\mathcal{N} \cap B(x,R)| \cdot \Vol(B(0,\delta/2)) \leq \Vol(B(0,R + \delta/2))\,,
    \end{equation*} 
    which implies the fact.
\end{proof}

We now build a net for possible tree embeddings.

\begin{lemma}\label{lem:tree-net-construction}
    For each $k \geq 1$ and $\gamma \in (0,1/4)$, there is a collection $\mathcal{C}_k$ of tree embeddings so for all $\{x_\omega\} \in \mathcal{C}_k$ we have $x_{0^k} = 0$ and with the following additional property.  For any tuple $\{y_{\omega}\}_{\omega \in T_k}$ so that $y_{0^k} = 0$ with the property $$\frac{1}{2} A \cdot 4^{d(\omega,\omega')} \leq |y_\omega - y_{\omega'}| \leq \frac{3}{2} A \cdot 4^{d(\omega,\omega')}$$
    there is some tuple $\{x_\omega\} \in \mathcal{C}_k$ so that $|x_\omega - y_\omega| \leq \gamma$ for all $\omega \in T_k$.  Further, the collection $\mathcal{C}_k$ satisfies \begin{equation*}
        |\mathcal{C}_k| \leq \exp\left(C_\gamma  d 2^k \right)\,.
    \end{equation*}
\end{lemma}
\begin{proof}
    We will use Fact \ref{fact:basic-net} with $\delta = \gamma$ to obtain a $\gamma$ net $\mathcal{N}$. 
    
    We will in fact define points $x_\omega$ for $\omega$ on the binary tree, rather than simply for $\omega$ on the leaves of the tree.  To begin with, for the root $\emptyset$ of the tree initialize $x_\emptyset = 0$.  For each binary word $\sigma$ of length strictly less than $k$, set $j = k - |\sigma|$ and iteratively define \begin{equation}\label{eq:tree-net-choice}
        x_{\sigma 0 } = x_\sigma\,,\quad x_{\sigma 1} \in \mathcal{N} \cap B(x_\sigma, \frac{3}{2} A 4^j + 2\gamma)
    \end{equation}
    and note that by Fact \ref{fact:basic-net} the number of choices for $x_{\sigma 1}$ is at most $(5 + 3 A 4^j / \gamma)^d \,.$  Because there are $2^{k - j}$ binary words of length $k - j$, the number of choices total is at most \begin{equation*}
        \prod_{j = 1}^k (5 + 3 A 4^j / \gamma)^{d 2^{k-j}} \leq \exp(C_\gamma d 2^k)
    \end{equation*} 
    for some $C_\gamma < \infty$ depending on $\gamma$ and $A$. 
    Let $\mathcal{C}_k$ denote the collection of all possible tree embeddings obtained in this fashion and note that by the above bound we have $|\mathcal{C}_k| \leq \exp\left(C_\gamma  d 2^k \right)$.

    To complete the proof, we need to show that for a tuple $y_\omega$ satisfying the hypotheses of the lemma there is an $x_\omega \in \mathcal{C}_k$ that approximates it.  Let $\pi : \R^d \to \mathcal{N}$ be a map that satisfies $\|z - \pi(z)\|_2 \leq \gamma$ as guaranteed by Fact \ref{fact:basic-net} and impose $\pi(0) = 0$.  For a string $\sigma$ of length at most $k$ define $a_\sigma = y_{\sigma 0^{k - |\sigma|}}$ and $x_\sigma = \pi(a_\sigma).$  Note that for a word $\sigma$ with $|\sigma| < k$ we have that $$\|a_\sigma - a_{\sigma1}\| \leq \frac{3}{2}  A 4^j\,,$$
    and so the rounded values $x_\sigma$ and $x_{\sigma 1}$ satisfy Eq.~\eqref{eq:tree-net-choice}, thus showing that the tuple $(x_\omega)_{\omega\in T_k}$ obtained in this way lies in the set constructed above.  Further, for a pair $\omega,\omega' \in T_k$ we have $$\|x_{\omega} - x_{\omega'}\|_2 \geq \frac{1}{2} A 4^j - 2 \gamma  \geq \frac{1}{4}  A 4^j\,,$$
    thus showing that $(x_\omega)_{\omega \in T_k}$ is a tree embedding and thus in $\mathcal{C}_k$.
\end{proof}

\subsection{Completing the proof}

To complete the proof, we now combine the net construction of Lemma \ref{lem:tree-net-construction} with the probability estimate Lemma \ref{lem:tree-esimate}.

\begin{proof}[Proof of Proposition \ref{prop:upper-bound}]
    Note that if $0$ lies in an unbounded component of $\open$, then for each $k$ there is a  path in $\open$ connecting $0$ to $\partial B(0, A 4^k)$.  By Lemmas \ref{lem:extract-tree} and \ref{lem:tree-net-construction}, this implies that there is some tree embedding $\{x_\omega\}_{\omega \in T_k} \in \mathcal{C}_k$ so that for all $\omega \in T_k$ we have $\dist(x_\omega, \open) \leq \gamma$.  Taking $\gamma = t/2$, this shows that the event $\bigcap_{\omega \in T_k} V_{x_\omega}$ holds.   Summing over choices of the tree embedding $\{x_\omega\}_{\omega} \in \mathcal{C}_k$ and applying the bound Lemma \ref{lem:tree-esimate} bounds 
    \begin{equation*}
        \theta_d(\Phi) \leq |\mathcal{C}_k| e^{-c_1 \Phi 2^k} \leq \exp\left(C_\gamma d 2^k - c_1 M d 2^k \right)\,.
    \end{equation*}
    Taking $M$ large as a function of $C_\gamma$ and $c_1$ gives the bound $$  \theta_d(\Phi) \leq e^{- c_1 M d 2^k / 2}$$
    for each $k$.  Taking $k \to \infty$ shows $\theta_d(\Phi) = 0$ as needed.
\end{proof}

\section{Lower bound}
\label{sec:lower}

We now prove the lower bound of Theorem \ref{thm:bounds}:
\begin{proposition}\label{prop:lower-bound}
    There is a universal constant $\alpha > 0$ so that for $\Phi = \alpha d$ and $d$ large enough we have $$\theta_d(\Phi) > 0  \,.$$
\end{proposition}

\begin{figure*}
    \centering
    \includegraphics[width=\linewidth]{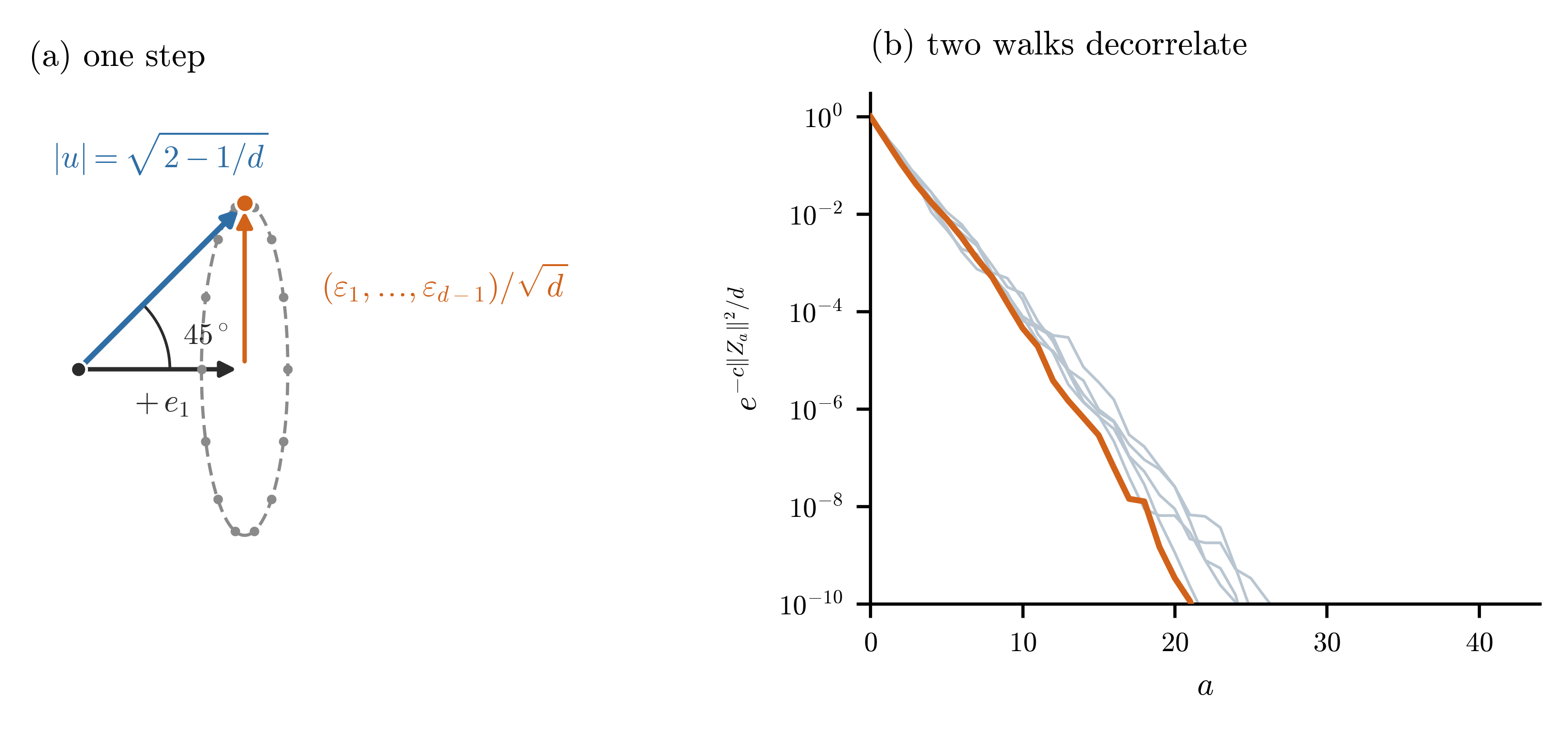}
    \caption{(a) Depiction of a single step $u = \bigl(1, \varepsilon_1/\sqrt{d}, \ldots,
    \varepsilon_{d-1}/\sqrt{d}\bigr)$, $\varepsilon_i = \pm 1$.  Each transverse
    coordinate moves  only by $1/\sqrt{d}$ and is individually negligible, but given that there are $d-1$ of them, the transverse increment (orange) has norm
    $\sqrt{(d-1)/d} \simeq 1$. In other words, every step advances one unit along $e_1$ and one
    unit transverse to it, making an angle of $45^\circ$ with $e_1$ and having
    length $|u| = \sqrt{2 - 1/d}$.  Its direction is one of the $2^{d-1}$
    corners of the transverse hypercube (dashed sphere, seen edge-on), all
    equally likely, so that $|\Gamma_k| = 2^{(d-1)k}$; this is the entropy the
    second-moment argument spends.  Accumulated over $k$ steps the same contrast
    makes a walk in $\Gamma_k$ ballistic along $e_1$ and diffusive transverse to
    it, so the region $U(\gamma)$ that must be kept empty of
    $\bX$ is a tube of length $k$ and transverse radius $\sim \sqrt{k} + R$.
    (b) Illustration of why two candidate walks barely interact.  The transverse difference $Z_a = S_a - T_{a+j}$ of two independent walks is itself a random walk on $\Z^{d-1}$, so the weight $e^{-c\|Z_a\|^2/d}$ that controls $\Vol(U(\gamma) \cap U(\gamma'))$ collapses within $O(1)$ steps; several realizations are shown for $d = 200$, one highlighted.  Summability of this weight is the content of Lemma~\ref{lem:random-walk-functional}, which bounds $\E[Z_k^2]$ and hence yields Eq.~\eqref{eq:Z_k-second-moment-bound}.}
    \label{fig:lower-bound}
\end{figure*}

We will define a set of walks $\Gamma_k$ of length $k$ that will serve as candidate paths.  First define $\mathcal{S}_k = \{ (S_j)_{j = 0}^k \subset \Z : S_0 = 0, |S_i - S_{i+1}| = 1\}$ be the set of simple walks on $\Z$ starting at $0$.

We then define
\begin{align*}
    \Gamma_k =& \Bigg\{ \bigg(\bigg(j, \frac{S_j^{(1)}}{\sqrt{d}},\frac{S_j^{(2)}}{\sqrt{d}}, \ldots \frac{S_j^{(d-1)}}{\sqrt{d}} \bigg)\bigg)_{j = 0}^k  &\\
    &\quad : (S_{j}^{(i)})_{j = 0}^k \in \mathcal{S}_k  \text{ all }i \in [d-1] \Bigg\} \,.
\end{align*}

In other words, $\Gamma_k$ consists of walks that always increase by $1$ in the first coordinate, and in each other coordinate takes a step of $\pm 1/\sqrt{d}$ (see Figure \ref{fig:lower-bound}).  Define $R= \sqrt{r^2 + 1}$ and note that if for a walk $\gamma \in \Gamma_k$ we have that $\bX \cap \left(\bigcup_{j = 0}^k B(\gamma_j,R) \right) = \emptyset$ then the continuous path given by connected line segments along adjacent vertices of $\gamma$ lies in $\open$.  With this in mind, for a path $\gamma \in \Gamma_k$ define $$U(\gamma) = \bigcup_{j = 0}^k B(\gamma_j,R)\quad\text{ and } \quad H(\gamma) = \Vol(U(\gamma))\,.$$

For a path $\gamma \in \Gamma_k$ define the event $$\mathcal{D}(\gamma) = \{\bX \cap U(\gamma) = \emptyset\}\,. $$
We now define a weighted count of open paths in $\Gamma_k$ by \begin{equation}\label{eq:Z_k-def}
    Z_k = \frac{1}{|\Gamma_k|}\sum_{\gamma \in \Gamma_k} e^{\Phi H(\gamma)} \mathbf{1}_{\mathcal{D}(\gamma)} \,.
\end{equation}
By linearity of expectation we have that $\E Z_k = 1$.  We first note that lower bounding the event $\{Z_k > 0\}$ is sufficient to show $\theta_d(\Phi) > 0\,.$

\begin{lemma}\label{lem:Z_k-witness}
    We have $\theta_d(\Phi) \geq \lim_{k \to \infty} \P(Z_k > 0)\,.$
\end{lemma}
\begin{proof}
    Note first that the events $\{Z_k > 0\}$ are decreasing in $k$ and so the limit in the lemma exists.  Further, note that on the event $\{Z_k > 0\}$, there is a continuous path from $0$ to $\partial B(0,k)$ contained in $\open$. By K\"onig's lemma, if for each $k$ there is a continuous path from $0$ to $\partial B(0,k)$ contained in $\open$ then $0$ lies in an unbounded component in $\open$. Altogether this shows
    \begin{equation*}
        \theta_d(\Phi) \geq \lim_{k \to \infty} \P(Z_k > 0)
    \end{equation*}
    which completes the proof.
\end{proof}

We will lower bound $Z_k$ using the second moment method.  To this end, if we let $\gamma$ and $\gamma'$ be chosen from $\Gamma_k$ independently and uniformly at random, note that \begin{align}
    \E[Z_k^2] &= \frac{1}{|\Gamma_k|^2}\sum_{\gamma,\gamma' \in \Gamma_k}e^{\Phi H(\gamma) + \Phi H(\gamma') } \P(\mathcal{D}(\gamma) \cap \mathcal{D}(\gamma')) \nonumber \\
    &= \frac{1}{|\Gamma_k|^2}\sum_{\gamma,\gamma' \in \Gamma_k} e^{\Phi \Vol(U(\gamma) \cap U(\gamma'))} \nonumber  \\
    &= \E_{\gamma,\gamma'} e^{\Phi \Vol(U(\gamma) \cap U(\gamma'))}\,.\label{eq:second-moment}
\end{align}

If we write $\gamma = \left(\left(j, \frac{S_j^{(1)}}{\sqrt{d}},\frac{S_j^{(2)}}{\sqrt{d}}, \ldots \frac{S_j^{(d-1)}}{\sqrt{d}} \right)\right)_{j = 0}^k$ and $\gamma' = \left(\left(j, \frac{T_j^{(1)}}{\sqrt{d}},\frac{T_j^{(2)}}{\sqrt{d}}, \ldots \frac{T_j^{(d-1)}}{\sqrt{d}} \right)\right)_{j = 0}^k$ then each $S_j^{(i)}$ and $T_j^{(i)}$ is a simple random walk on $\Z$ and they are mutually independent. 
\begin{lemma}\label{lem:two-replica-by-random-walk}
In the notation above there are universal constants $c,C > 0$ such that we have
\begin{align*}
    \Vol&(U(\gamma) \cap U(\gamma')) \\
    &\leq C \sum_{a,b = 0}^k e^{-c (a - b)^2} \exp\left( -\frac{c}{d}\sum_{i = 1}^{d-1}(S_a^{(i)} - T_b^{(i)})^2 \right)\,.
\end{align*}
\end{lemma}
\begin{proof}
    This is because \begin{equation*}
        \Vol(U(\gamma) \cap U(\gamma')) \leq \sum_{a,b = 0}^k \Vol(B(\gamma_a,R) \cap B(\gamma_b',R))\,,
    \end{equation*}
    and we use Fact \ref{fact:overlap-gaussian-decay} after noting that $(R/r)^d = O(1)\,.$
\end{proof}

Using the representation in Eq.~\eqref{eq:second-moment} together with Lemma \ref{lem:two-replica-by-random-walk}, we see that bounding the second moment of $Z_k$ can be reduced to bounding the exponential moment of a certain random walk functional.  

\begin{lemma}\label{lem:random-walk-functional}
    Let $c > 0$ be given.  There is a constant $\beta > 0$ so that for all $d$ sufficiently large if we define 
    $$Y_k =  \sum_{a,b = 0}^k e^{-c (a - b)^2} \exp\left( -\frac{c}{d}\sum_{i = 1}^{d-1}(S_a^{(i)} - T_b^{(i)})^2 \right)$$
    for i.i.d.\ simple random walk $S_a^{(i)}$ and $T_b^{(i)}$ on $\Z$ then we have 
    $\E e^{\beta dY_k} \leq 2^{d}$ uniformly in $k$.
\end{lemma}

We defer the proof to the next subsection, where the upper bound will follow from reducing to the $d = 3$ case.

With Lemma \ref{lem:random-walk-functional} in tow we may now complete the lower bound. 

\begin{proof}[Proof of Proposition \ref{prop:lower-bound}]
    By Lemma \ref{lem:Z_k-witness} it is sufficient to show a uniform lower bound on $\P(Z_k > 0)$.  By the Paley--Zygmund inequality we have \begin{equation}\label{eq:paley-zygmund}
        \P(Z_k > 0) \geq \frac{(\E Z_k)^2}{\E Z_k^2} = (\E Z_k^2)^{-1} \,.
    \end{equation}
    By Eq.~\eqref{eq:second-moment} along with the bound in Lemma \ref{lem:two-replica-by-random-walk} and using the notation of Lemma \ref{lem:random-walk-functional} we see \begin{equation}\label{eq:Z_k-second-moment-bound}
        \E Z_k^2 \leq \E \exp(C \Phi Y_k) = \E \exp\left(C \alpha d Y_k \right)\,.
    \end{equation}
    Taking $\alpha$ small enough as a function of $C$ and $\beta$ from Lemma \ref{lem:random-walk-functional} shows $\E Z_k^2 \leq 2^d$.  Combining this estimate with Eq.~\eqref{eq:paley-zygmund} and Lemma \ref{lem:Z_k-witness} shows \begin{equation*}
        \theta_d(\Phi) \geq 2^{-d},
    \end{equation*}
    thus completing the proof.
\end{proof}

\subsection{Bounding the random walk functional}

To begin with, we prove a bound on the case $d=3$, which will provide the majority of the work.  We start with a basic bound that follows from transience of simple random walk on $\Z^3$.
\begin{lemma}\label{lem:3D-SRM-gaussian-sum}
    Let $M_n$ be the random walk on $\Z^3$ whose increments are given by a difference of two independent uniform vectors in $\{-1,1\}^3$.  Let $\E_x$ denote the expectation under the condition $M_0 = x$.  Then for each $c > 0$ we have \begin{equation*}
        \sup_{x} \E_x \sum_{n \geq 0} \exp(-c \|M_n\|_2^2) < \infty \,.
    \end{equation*}
\end{lemma}
\begin{proof}
    By transience of $M_n$, if we let $G(x,y)$ denote the Green's function of the random walk with the given increments  we have $G(x,y) \leq G(0,0) < \infty$.  Linearity of expectation shows
    \begin{align*}
        \E_x \sum_{n \geq 0} \exp(-c \|M_n\|_2^2)& = \sum_{y \in \Z^3} G(x,y) \exp(-c \|y\|_2^2) \\
        &\leq G(0,0) \sum_{y \in \Z^3}  \exp(-c \|y\|_2^2) \\
        &< \infty 
    \end{align*}
    completing the estimate.
\end{proof}

This uniform estimate translates to a simple upper bound on the corresponding exponential moment

\begin{corollary}\label{cor:exponential-moment-3D}
    For each $c > 0$, there is a $\beta = \beta(c) > 0$ so that \begin{equation*}
        \sup_x \E_x \exp\left(\beta \sum_{n \geq 0} \exp\left(-c \|M_n\|_2^2 \right) \right) \leq 2\,.
    \end{equation*}
\end{corollary}
\begin{proof}
    Let \begin{equation*}
        \Lambda = \sup_{x} \E_x \sum_{n \geq 0} \exp(-c \|M_n\|_2^2) < \infty\,,
    \end{equation*}
    where the finiteness is guaranteed by Lemma \ref{lem:3D-SRM-gaussian-sum}.  Note that \begin{align*}
        \frac{1}{k!}\sup_x \E_x \left( \sum_{n \geq 0} \exp\left(-c \|M_n\|_2^2 \right)\right)^k \leq \Lambda^k
    \end{align*}
    by the Markov property.  This implies that 
    \begin{align*} 
    \sup_x \E_x \exp\left(\beta \sum_{n \geq 0} \exp\left(-c \|M_n\|_2^2 \right) \right) &\leq \sum_{k \geq 0} \Lambda^k \beta^k\\
    &\leq (1 - \Lambda \beta)^{-1},\end{align*}
    where the last bound is for $|\Lambda \beta| < 1$.  Taking $\beta$ small enough completes the proof.
\end{proof}

\begin{proof}[Proof of Lemma \ref{lem:random-walk-functional}]
    Define  $I_j = \{a \in \Z : 0 \leq a \leq k, 0 \leq a + j \leq k\}$
    \begin{align*}
    Y_k &\leq \sum_{j = -k}^k e^{-c j^2} \sum_{a \in I_j} \exp\left( -\frac{c}{d}\sum_{i = 1}^{d-1}(S_a^{(i)} - T_{a+j}^{(i)})^2 \right)\\
    &=: \sum_{j = -k}^k e^{-c j^2} W_j \,. 
    \end{align*}
    Writing $\sum_{j \in \Z} e^{-c j^2} = \alpha_c$, Jensen's inequality implies \begin{equation*}
        \E e^{\beta d Y_k} \leq \sup_{|j| \leq k} \E e^{\beta d \alpha_c W_j}\,.
    \end{equation*}
    For a fixed $j$, note that the vectors $(S_a^{(i)} - T_{a+j}^{(i)})_{i = 1}^{d-1}$ as $a$ varies form a random walk in $\Z^{d-1}$ whose increments are the difference of independent copies of the uniform vector on $\{-1,1\}^{d-1}.$  In particular, we can write it as $(z_a)_{a = 0}^k$ where $z_a \in \Z^{d-1}$ is the random walk with increments as previously described, and so we are interested in bounding the exponential moment of \begin{equation*}
        \sum_{a = 0}^k \exp\left(-\frac{c}{d}\|z_a\|_2^2\right)\,. 
    \end{equation*}
    
    Set $m = \lfloor (d - 1)/3 \rfloor$ and define $D = 3m$.  Group the first $D$ coordinates of $z_a$ into $m$ blocks of $3$ and label these blocks as $\xi_a^{(1)},\ldots,\xi_a^{(m)}$ where each $\xi_a^{(j)} \in \Z^3$.  In particular, we have \begin{align*}
        \sum_{a = 0}^k \exp\left(-\frac{c}{d}\|z_a\|_2^2\right) &\leq \sum_{a = 0}^k \exp\left(- \frac{c}{d} \sum_{\ell = 1}^m \|\xi_a^{(\ell)}\|^2_2 \right) \\
        &\leq \frac{1}{m}\sum_{a = 0}^k \sum_{\ell = 1}^m  \exp\left(- \frac{cm}{d} \|\xi_a^{(\ell)}\|^2_2 \right)
    \end{align*}
    where the last inequality is via the AM-GM inequality.  Applying Corollary \ref{cor:exponential-moment-3D} completes the proof.
\end{proof}

\section{Conclusion}
\label{sec:conclusion}

\begin{figure}
    \centering
    \includegraphics[width=\linewidth]{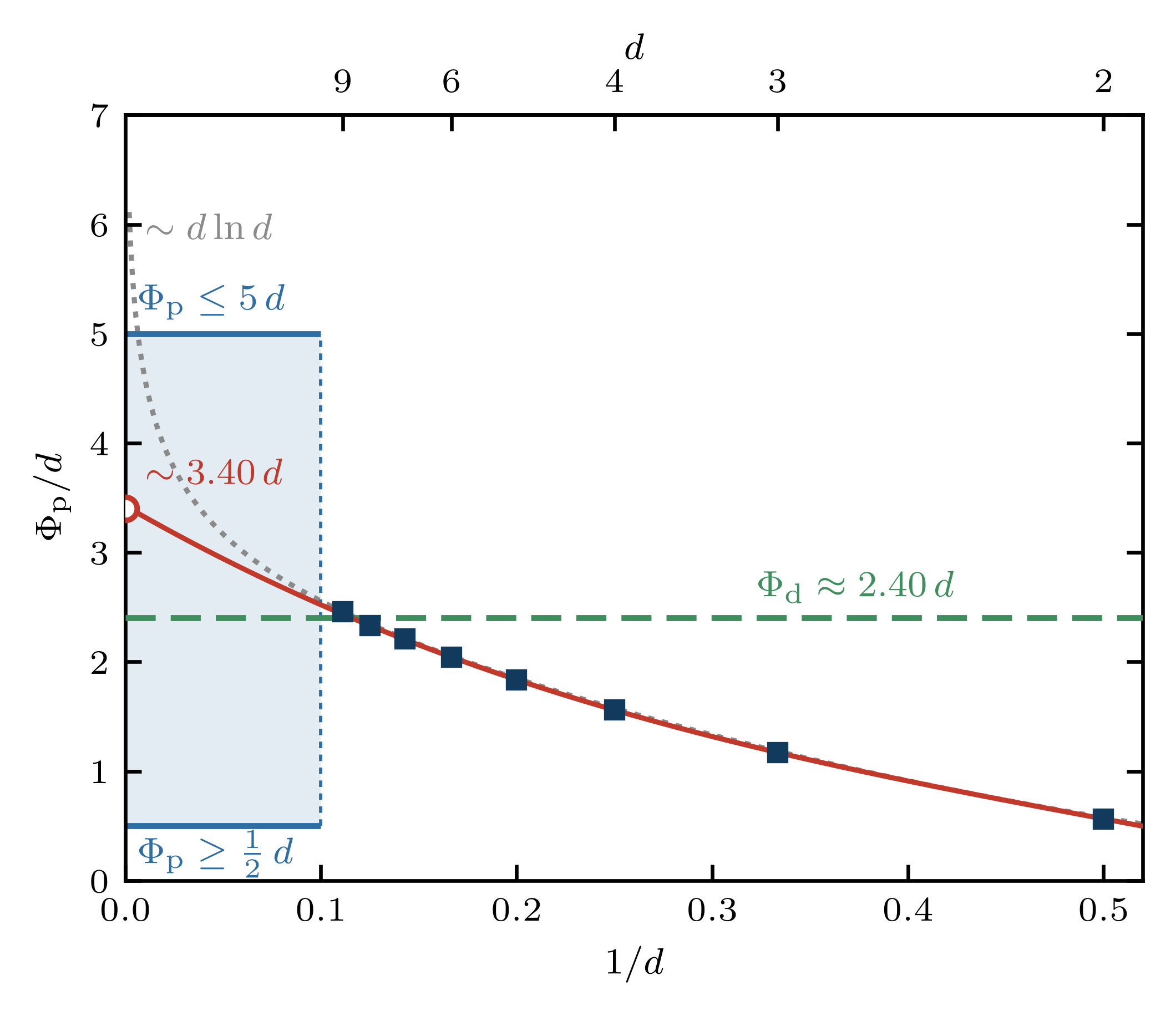}
    \caption{Dimensional scaling of the void percolation threshold, including numerical results of Ref.~\cite{Charbonneau2021} (squares) and two     extrapolations: (i) the cubic in $1/d$ [Eq.~(26) of \cite{Charbonneau2021}] (red line)  approaching a constant $\Phi_{\rm p}/d \to 3.425(50)$ (open circle), and the alternative saturating form $\Phi_{\rm p} \sim d\ln d$ [Eq.~(27) of \cite{Charbonneau2021}] (gray dashed line), which diverges on this scale.  The numerics alone cannot distinguish them: the two curves separate only well beyond $d = 9$. The window established in this work, $\tfrac{1}{2}\,d \leq \Phi_{\rm p} \leq 5\,d$ (shaded band) settles the ambiguity, excluding the $d \ln d$ scaling and confirming $\Phi_{\rm p} = \Theta(d)$. The mean-field dynamical transition $\Phi_{\rm d} = 2.40\,d$~\cite{Biroli2021} (dashed green line) is crossed in $d = 9$, so that the window $[\Phi_{\rm d}, \Phi_{\rm p}]$ in which activated escape is expected to control the approach to percolation opens slowly with dimension for $d \gtrsim 9$.}
    \label{fig:summary}
\end{figure}
The key result of this work, Theorem~\ref{thm:bounds}, markedly improves on previously known and conjectured bounds on the void percolation threshold of the RLG.

A first immediate consequence of this analysis is that the proposed $d\ln d$ scaling of this threshold is excluded.  As detailed in Fig.~\ref{fig:summary}, the numerical thresholds of
Ref.~\cite{Charbonneau2021} for $d = 2$ to $9$ admit two extrapolations that
the data cannot separate: a cubic in $1/d$ approaching a constant
$\Phi_{\rm p}/d \to 3.40(5)$, and a saturating form $\Phi_{\rm p} \sim d \ln d$
that diverges on those axes.  The two curves are indistinguishable over the
whole simulated range and part company only well beyond $d = 9$. No
foreseeable extension of the numerics could have decided between them. 

The results further motivate the quest to understand instantons in the RLG and in the broader class of RFOT models.  Although the range of numerical prefactors in Theorem~\ref{thm:bounds} remains too broad to confirm it, numerical results suggest that cage escapes above $\Phi_\mathrm{d}$ do exist in sufficiently high $d$, but persist only up to a finite $\Phi/d\approx 3.4$. Because the mean-field dynamical transition is also linear, $\Phi_{\rm d}/d \approx 2.40$~\cite{Biroli2021}, Theorem~\ref{thm:bounds}
therefore likely fixes the fate of the interval $[\Phi_{\rm d}, \Phi_{\rm p}]$ over which activated processes are expected to control the approach to arrest in sufficiently high $d$. It is important to note that this scenario is remarkably different from what has been suggested by other mechanisms thought to drive activated processes, such as the caging complexity~\cite{parisi2019} and the cage stability relative to hopping~\cite{ikeda2026}, both of which scale as $\Phi/d\sim \ln d$. These discrepancies highlight the limited extent to which instantonic processes in this class of systems are understood, and further motivate an effort dedicated to their study.

Given the ongoing struggles of theoretical physics to address this key problem, ample room also remains for insights from a probabilistic approach. In this respect, several questions follow directly from the current work.  First, one might reasonably wonder whether the constants $c$ and $C$ can be brought together. Even a rigorous proof that $\Phi_{\mathrm{p}}(d) / d$ converges to a limiting constant---without identifying the said constant---would be interesting and appears to require new ideas.  This problem may be quite mathematically challenging, and bears a certain resemblance to the problem of determining that the optimal sphere packing density of Euclidean space has an exponential rate; the current state of the art has bounds that differ by an exponential factor: all lower bounds are of the form $(1/2 + o(1))^d$ \cite{klartag2026lattice} while the best upper bound is $(\sqrt{\frac{e}{2\pi}} + o(1))^d$ \cite{OpenAI2026TenAdvances}. Second, the sub-leading structure offers an interesting target. The finite-$d$ correction to $\Phi_{\mathrm{p}}(d)/d$, approximately fitted to $-10/d$ (Fig.~\ref{fig:summary}), is large enough to be significant in the numerically accessible regime. Any rigorous access to it would therefore test the extrapolation even more stringently than the leading order does. Third, one might consider the underlying geometric question: what do the vacant cluster look like at threshold on the scale at which both proofs operate, and how does that geometry relate to the cages of the mean-field description? 

Given the rich and interesting set of questions that remain open, we trust that the linear scaling of the void percolation threshold offers a broadly usable starting point for both physics and probability.

\paragraph*{Data Availability Statement}
Data relevant to this work have been archived and can be accessed at the Duke Digital Repository~\cite{RDR}. [Data will be made openly available upon manuscript acceptance, but a DOI is not available at the time of submission.]

\paragraph*{AI declaration}
GPT 5.6 suggested the notion of a tree embedding for the upper bound along with the proof idea for Lemma \ref{lem:random-walk-functional}.  After the authors wrote the main body of text, GPT 6 was used to optimize the constants and come up with the proof in Appendix \ref{sec:appendix}, including the writing the code for all programs that verify numerical checks.  GPT 6 was used for proofreading and copy-editing.
M.M.\ acknowledges OpenAI’s ChatGPT for Academic Researchers program for providing complimentary access to ChatGPT.

\paragraph*{Acknowledgment}
P.C.\ acknowledges various stimulating conversations with Giampaolo Folena and Francesco Zamponi. M.M.\ is supported in part by NSF Grants DMS-2618808 and DMS-2643929. 
The authors acknowledge support from the National Institute for Theory and Mathematics in Biology (NITMB) through the National Science Foundation (grant number DMS-2235451) and the Simons Foundation (grant number MP-TMPS-00005320).  This work began at NITMB during a workshop titled ``Spin Glass Theory and High-Dimensional Landscapes in Biology'' in August 2026.

\appendix

\section{Explicit constants} \label{sec:appendix}

In this appendix we indicate how to adapt the proof of Theorem \ref{thm:bounds} in order to yield explicit constants of $1/2$ and $5$ for $c$ and $C$, respectively.

\subsection{Upper bound}

Set \begin{equation}
    h_d = d/ r^2, \quad u =v = 3/8, \quad t = u/\sqrt{h_d}, \quad \Phi = 5d\,.
\end{equation}

We first prove an upgraded version of Lemma \ref{lem:one-point-estimate}.  For this, let $N$ denote the matrix with $k$ rows, each of which is an independent unit vector in $\R^d$.  Define $H = N N^T$.  For weights $b_i \geq 0$ and $k < d$, define $C_{d,k}(b) = (1 - \sum_i b_i/r)_+^{d- k}$; for $d = k$, define $C_{d,d}(b) = 1$.  Define the sum \begin{equation*}
    B_d =\sum_{k = 0}^d \frac{\Phi^k}{k!}\left(\frac{d}{r} \right)^k \E \left[\det H \int_{|N^T b| \leq t,b_i \geq 0} C_{d,k}(b)\,db \right]\,,
\end{equation*}
where we interpret the $k = 0$ summand as $1$.  We can now state: 
\begin{lemma}\label{lem:app-one-point} We have
    $$\P(\open \cap B(0,t) \neq \emptyset) \leq e^{-5d} B_d\,.$$
\end{lemma}
\begin{proof}
    Put $\mathcal W=\{x:\dist(x,\bX)\geq r\}$, a closed set containing $\open$, and label $\cE = \{\mathcal W \cap B(0,t) \neq \emptyset\}$. It suffices to bound $\P(\cE)$. On the event $\cE\cap\{0\notin\mathcal W\}$
    let $x\in\mathcal W$ minimize $\|x\|_2^2$, which exists by closedness and satisfies $\|x\|_2\leq t$.  Let $z_1,\ldots,z_k$ be the set of points for which $\|x - z_i\|_2 = r$ and set $N_i = (x - z_i)/r$.  We note that $k \geq 1$ as otherwise $x$ would lie in the interior of $\mathcal W$ and could thus be moved closer to the origin.  We work on the probability-one general-position event on which at most $d$ radius-$r$ sphere boundaries meet at any point and the normals of every such intersection are linearly independent and so may assume without loss of generality that $H=NN^T$ is nonsingular. 

    For simplicity label $\cE = \{\open \cap B(0,t) \neq \emptyset \}.$  On the event $\cE \cap \{0 \notin \open\}$ 
    let $x \in \open$ minimize $\|x\|_2^2$, i.e.\ be a point on $\open$ closest to the origin.  Let $z_1,\ldots,z_k$ be the set of points for which $\|x - z_i\|_2 = r$, and set $N_i = (x - z_i)/r$.  We note that $k \geq 1$ as otherwise $x$ would lie in the interior of $\open$ and could therefore be moved closer to the origin.  Further, with probability $1$, for every point in $\R^d$ there are at most $d$ spheres placed whose boundaries contain that point, and so we will always work on this event of full probability.  Setting $H = N N^T$, we in particular also have that $H$ is nonsingular almost surely. 

    \begin{claim}\label{claim:first-order}
    There are non-negative weights $b_i \geq 0$ so that $x = \sum_{i = 1}^k b_i N_i$.
    \end{claim}
    \begin{proof}[Proof of Claim \ref{claim:first-order}]
        Given a vector $v$ satisfying $\langle v, N_i \rangle \geq 0$ for all $i$, we have that for $\eps$ small enough that $x + \eps v \in \mathcal{W}$.  In particular by minimality of $x$ we must have that $\langle x, v \rangle \geq 0$.  By duality this implies that there are some $b_i \geq 0$ so that $x = N^t b.$
    \end{proof}

    We now observe a bound on the sum of the weights $b_i$ from the previous claim.

    \begin{claim}\label{claim:second-order}
        Whenever $k < d$ we have $\beta = \sum_{i} b_i \leq r$.
    \end{claim}
    \begin{proof}[Proof of Claim \ref{claim:second-order}]
        The spheres of radius $r$ centered at $z_i$ form a $d - k$ dimensional smooth manifold with tangent space $T = \{v : \langle N_i,v\rangle = 0 \text{ for all }i\}$.  Choose a non-zero $v \in T$---which is always possible since $k < d$---and so $v$ is thus the velocity of a smooth curve $\gamma$ in this intersection with $\gamma(0) = x$.  In a neighborhood of $x$ this curve lies in $\mathcal{W}$.  Differentiating $\|\gamma(s) - z_i\|_2^2 = r^2$ twice gives $\langle N_i, \gamma''(0)\rangle  = -\|v\|^2_2 / r$.  Since $x = \sum_i b_i N_i$, minimality of $x$ gives \begin{equation*}
            0 \leq \|v\|_2^2 + x \cdot \gamma''(0) = (1 - \beta/r)\|v\|_2^2\,,
        \end{equation*}
        thus completing the proof.
    \end{proof}

    In order to convert this result into a probabilistic bound, we need to understand the Jacobian of the map taking the set of vectors $N_i$ and weights $b$ into the points $z_i$, which are placed according to the Poisson point process.  In particular, for $k \leq d$, $N_i \in \mathbb{S}^{d-1}$, and weights $b_i \geq 0$, set $x = N^T b$ and define $$F(N,b) = (x - rN_1,\ldots,x - rN_k)\,.$$
    \begin{claim}\label{claim:jacobian}
        For $\sum_i b_i  \leq r$ and $k < d$, the Jacobian of $F$ is $J_F = r^{k(d-1)} (\det H) C_{d,k}\,.$ For $k = d$, the same formula holds with $C_{d,d} = 1$, without the restriction on $\sum_i b_i$. 
    \end{claim}
    \begin{proof}
        By homogeneity it is sufficient to assume $r = 1$.  For $N_i \in \R^d$ and $x \in \R^d$ if we define $z_i = x - N_i$, $s_i = (\|N_i\|_2^2 - 1)/2$, and $c = x - N^T b$, then we can consider $$G:(N_1,\ldots,N_k,b,x) \mapsto (z_1,\ldots,z_k,s,c)\,,$$
        and note that $s = 0$ and $c = 0$ corresponds to our original parameterization. At $s = c = 0$, use local orthonormal spherical coordinates together with the transverse coordinates $s,c$.  The change into these coordinates has Jacobian $1$, and in these coordinates the differential $D G$ is block triangular with $J_F = | \det DG |$.  By direct computation we have \begin{equation*}
            |\det DG | = \left|\det\begin{bmatrix} 
            0 & N \\ -N^T & (1 - \beta)I_d\end{bmatrix} \right| = (\det H )|1 - \beta|^{d - k}.
        \end{equation*}
        By Claim \ref{claim:second-order} we have $\sum_i b_i \leq r$ whenever $k < d$, thus completing the proof.  
    \end{proof}

    We may sum over choices of $k$ with the $k = 0$ choice corresponding to $0 \in \mathcal{W}$.  Apply the area formula counting all preimages of $F$, with product surface measure for the normals and Lebesgue measure for $b$, followed by the Mecke formula. The remaining process must avoid the ball $B(x,r)$ which holds with probability $e^{-\Phi}$. This gives
    \begin{align*}
        \P(\mathcal{E}) &\leq \sum_{k = 0}^d e^{-\Phi} \frac{\Phi^k}{k!} \left(r^{d-1} \Vol_{d-1}(\mathbb{S}^{d-1}) \right)^k  \\
        &\quad \times \E \left[\det H \int_{|N^T b|
        \leq t,b_i \geq 0} C_{d,k}(b)\,db \right]\,.
    \end{align*}
    Noting that $r^{d-1}\Vol_{d-1}(\mathbb{S}^{d-1}) = d/r$ completes the proof.
\end{proof}

We then bound the exponential growth rate of $B_d$. Define \begin{align*}
    G_0 &= \eta_0 + \log \frac{5}{8(1 - \eta_0)},\quad \eta_0 = \frac{\sqrt{21} - 3}{2} \\
    G_1 &= \eta_1 + \log\frac{5}{8 \sqrt{1 - \eta_1}}, \quad 2\eta_1^3 = 9\pi(1 - \eta_1)
\end{align*}
where $\eta_1 \in (0,1)$ and $G = \max(G_0,G_1)$.  \begin{lemma}\label{lem:app-B-exp-rate}
We have     $\limsup_{d \to \infty} \frac{1}{d} \log B_d \leq G.$
\end{lemma}
\begin{proof}
    We split the integral at $\sum b_i = vr$ where we recall $v = 3/8$.  Note that $\E \det H = (d)_k / d^k$ and so the integral over $\sum b_i < vr$ may be bound by $(5d)^k \P(\mathrm{Bin}(d,v) \geq k) / k!$.  For the latter, bound $C_{d,k} \leq ( 1 - v)^{d - k}$ and note that the remaining integral gives $2^{-k} \kappa_k t^k \E \sqrt{\det H}$ after averaging over sign changes of the normals.  Direct computation gives \begin{equation*}
        \mu_{d,k} := \E \sqrt{\det H} = \left[\frac{\Gamma(d/2)}{\Gamma((d+1)/2)} \right]^k \frac{\Gamma((d+1)/2)}{\Gamma((d- k + 1)/2)}\,,
    \end{equation*}
    thus giving the bound \begin{align*}
        B_d &\leq 1 + \sum_{k = 1}^{d-1} \frac{(5d)^k}{k!} \P(\mathrm{Bin}(d,v) \geq k) \\
        &\quad + \sum_{k = 1}^d \frac{(5d)^k}{k!}(1 - v)^{d-k}  \kappa_k \left(\frac{u\sqrt{d}}{2} \right)^k \mu_{d,k}\,.
    \end{align*}
    We therefore need to compute the exponential rates of each term.  By Stirling's formula and the binomial Chernoff bound, if we write $\alpha = k/d$ then the exponential rate of the $k$th term of each sum is \begin{align*}
        I_0(\alpha) &= \alpha \log(5e /\alpha) - \mathbf{1}_{\alpha > v} D(\alpha||v) \\
        I_1(\alpha) &= \alpha \log(5u) + \alpha + \frac{\alpha}{2}\log(\pi/2) - \frac{3\alpha}{2} \log \alpha \\
        &\quad -\frac{1-\alpha}{2}\log(1 - \alpha) + (1 - \alpha) \log(1 - v).
    \end{align*}
    By taking a first derivative, $I_0$ is maximized at $\eta_0$ achieving value $G_0$ while $I_1$ is maximized at $\eta_1$ achieving value $G_1$.
\end{proof}

We now need a more quantitative version of Lemma \ref{lem:tree-esimate}.  For distinct points $z_1,\ldots,z_n \in \R^d$ let $a_{i,j} = \sqrt{h_d} |z_i - z_j|/2$, $F_d(a) = \P(\sqrt{d} V_1 \geq a)$ for $V$ uniform in the unit ball and $f_d = -F_d'$. Define \begin{equation*}
    E_{i,d} = \sum_{j \neq i} F_d(a_{i,j}), \quad P_{i,d} = \sum_{j \neq i} f_d(a_{i,j} - u)\,.
\end{equation*}

\begin{lemma}\label{lem:app-separation-indep}
    For all distinct points $z_1,\ldots,z_n$ with $a_{i,j} > u$ for $i \neq j$, if $E_{i,d} \leq E_d$ and $P_{i,d} \leq P_d$ for all $i$ then \begin{align*}
        \P&(\open \cap B(z_i,t) \neq \emptyset \text{ for all }i) \\
        &\leq \exp\left(-dn[5 - G - 5E_d - 225 P_d^2/128 - o(1)] \right) 
    \end{align*}
    where the $o(1)$ term depends only on $d$.
\end{lemma}
\begin{proof}
    Consider the Voronoi diagram associated to the points $z_1,\ldots,z_n$ and let $C_1,\ldots,C_n$ denote the corresponding Voronoi cells.  Note that the Poisson point processes in each $C_j$ are mutually independent.  Writing $x = z_i + \tau t U$ for some $\tau \in [0,1]$ and $\|U\|_2 = 1$ we see \begin{equation*}
        \Vol_d(B(x,r) \setminus C_i) \leq E_{i,d} + u \sum_{j \neq i} f_d(a_{i,j} - u) |\langle U,v_{i,j} \rangle|\,,
    \end{equation*}
    where $v_{i,j} = (z_j - z_i)/ \|z_j - z_i\|_2$.  Writing $p_i = \P(\exists~x \in B(z_i,t) : B(x,r) \cap C_i \cap \bX = \emptyset)$, we repeat the positive candidate integral in Lemma~\ref{lem:app-one-point}, discarding the restrictions that the active centers lie in $C_i$ and inserting the loss factor $e^{5d\Vol(B(x,r)\setminus C_i)}$. Conditional on the candidate radius, rotational invariance makes its direction $U$ uniform. We may therefore apply Jensen's inequality inside this integral with weights $\alpha_j \propto f_d(a_{i,j} - u)$ to bound 
    $$p_i \leq e^{-5d}B_d \cdot \left(e^{5d E_{i,d}} \sum_j \alpha_j \E_U e^{5du P_{i,d} |\langle U,v_{i,j}\rangle|} \right)\,,$$ where $U$ is uniform on $\mathbb{S}^{d-1}.$  Using the bound $\E e^{s|U_1|} \leq 2 e^{s^2 / (2d)}$ shows $$p_i \leq 2e^{-5d}B_d \exp\left(d \left(5 E_{i,d} + \frac{225}{128}P_{i,d}^2 \right) \right)\,. $$
    Multiplying over $i$ and using Lemma \ref{lem:app-B-exp-rate} completes the proof.
\end{proof}

We now show how to extract a tree in a similar fashion as Lemma \ref{lem:extract-tree}.  The main difference is that our tree has $32$ children per vertex as opposed to $2$.  Set $R_j = \frac{7(32^j -1)}{31}$, $\ell_0 = 0$ and $\ell_j = 32R_j - 7j$.  \begin{lemma}\label{lem:app-tree-extraction}
    Given a continuous curve starting at $0$ and reaching distance $\ell_k/\sqrt{h_d}$, there are $N = 32^k$ points including $0$ indexed by the leaves of a tree with $32$ children per vertex.  Two leaves at genealogical height $j$ are separated by at least $7j/\sqrt{h_d}$ and each leaf has $31 \cdot 32^{j-1}$ partners at height $j$.  This forms a tree on these points, rooted at $0$ with $31\cdot 32^{k-j}$ edges of length $R_j/\sqrt{h_d}$.
\end{lemma}
\begin{proof}
    Rescale space to set $\sqrt{h_d} =1$.  The proof is identical to the proof of Lemma \ref{lem:extract-tree} using the identities $R_j = \ell_{j-1} + 7j$ and $\ell_j = 31 R_j + \ell_{j-1}$ for the inductive step.
\end{proof}

We now bound the quantities in Lemma \ref{lem:app-separation-indep} for points obtained from Lemma \ref{lem:app-tree-extraction}.  For fixed $J \geq 1$ and $\eps \in (0,u)$ define \begin{equation*}
    a_j^{J,\eps} = \begin{cases}
        7j/2 - \eps, \quad j \leq J, \\
        7j/2 - u, \quad j > J
    \end{cases}
\end{equation*}
and define \begin{align*}
    E_d &= \sum_{j \geq 1} 31 \cdot 32^{j-1} F_d(a_j^{J,\eps}) \\
    P_d &= \sum_{j \geq 1} 31 \cdot 32^{j-1} f_d(a_j^{J,\eps} - u)\,.
\end{align*}

We introduce an observable that will take the place of the union bound appearing in the proof of Proposition \ref{prop:upper-bound}.
\begin{lemma}\label{lem:app-UB-markov}
    For each $k$ there is a non-negative random variable $X_k$ so that with $n = 32^k$ we have the following properties: if an open curve connects $0$ to distance $\ell_k/\sqrt{h_d}$ then $X_k \geq e^{-dn(H_\ast + o(1))}$ for $$H_\ast = \sum_{j \geq 1} 31 \cdot 32^{-j} \log \frac{\sqrt{R_j^2 + 2u^2}}{u}\,,$$
    where the error is uniform in $k$ (with fixed $J$ and $\eps$).  We further have $$\E X_k \leq e^{-dn\left(5 - G - 5E_d - \frac{225}{128}P_d^2 - o(1)\right)}\,.$$
\end{lemma}
\begin{proof}
    The proof is similar to that of Lemma \ref{lem:tree-net-construction}, but instead of a discrete net we will define an auxiliary Gaussian probability distribution on the points obtained from Lemma \ref{lem:app-tree-extraction}.  Select $n = 32^k$ points using Lemma \ref{lem:app-tree-extraction}.  For an edge of the introduction tree in Lemma~\ref{lem:app-tree-extraction}, joining $v$ to its parent $p(v)$, let $j=j(v)$ be the height of the extraction step that introduced it, and define $\sigma_j^2 = \frac{R_j^2 + 2u^2}{dh_d}$ and $q_j$ the density of a centered Gaussian with covariance $\sigma_j^2 I_d$.  Fix the root position $z_0 = 0$, and recursively generate the remaining positions by $z_v = z_{p(v)} + G_v$, where $G_v \sim q_j$ are independent and $p(v)$ is the parent of $v$.  Conditional on its parent's position, a new point $z_v$ has density $q_{j(v)}(z_v - z_{p(v)})$, and so the density of all non-root positions is $$\rho_k(z) = \prod_{v \neq 0} q_{j(v)}(z_v - z_{p(v)}).$$  
    Fix the genealogical labels, and let $\mathcal T$ consist of configurations for which a pair of height $j$ satisfies $\sqrt{h_d}|z_v-z_w|\geq 7j-2\eps$ for $j\leq J$, and $\sqrt{h_d}|z_v-z_w|\geq7j-2u$ for $j>J$. With $z_0=0$ fixed and $dz$ integrating only the non-root positions, define
    \begin{equation*}
        X_k = \int_{\mathcal{T}} \rho_k(z) \prod_v \mathbf{1}_{\open \cap B(z_v,t) \neq \emptyset}\,dz\,.
    \end{equation*}
    Because $\rho_k$ is a density, Lemma \ref{lem:app-separation-indep} confirms the upper bound on $\E X_k$.  For the lower bound on $X_k$, use Lemma \ref{lem:app-tree-extraction} to extract a tree from the open curve and perturb each nonroot point in a radius $u/\sqrt{h_d} = t$ ball, restricting its projection onto each direction towards a partner of height at most $J$ to $[-eps/\sqrt{h_d}, \eps/\sqrt{h_d}]$, independently for the non-root points; keep the root fixed. Each retained set is centrally symmetric. There are at most $32^J-1$ slab restrictions per point, so its volume fraction is at least $1-(32^J-1)u^2/((d+2)\eps^2)=1-o(1)$. The perturbations have mean zero, preserve the defining separations of $\mathcal T$, and leave each original vacant point inside its test ball. The mean squared length of the edge is at most $(R_j^2 + 2u^2)/h_d$.  Jensen's inequality and Stirling's formula then bound the negative logarithm of the retained Gaussian mass by $d\log(\sqrt{R_j^2+2u^2}/u)+o(d)$ per introduction edge. The error depends only on $d,J,\eps$, uniformly in the edge height. Summing over the $31\cdot32^{k-j}$ edges at each height proves the lower bound with error $o(dn)$ uniformly in $k$.
\end{proof}

We are nearly ready to put the pieces together for the proof.  Before this, we need one last numerical check.  Let $Q(t) = \P(\mathcal{N}(0,1) \geq t)$ and $\phi$ denote the density of a standard Gaussian.  First let $d \to \infty$ with $J$ and $\eps$ fixed; then let $\eps \to 0$ and $J \to \infty$.  In these successive limits the quantities $E_d$ and $P_d$ converge to \begin{align*}
    E_\ast &= \sum_{j \geq 1} 31 \cdot 32^{j-1} Q(7j/2) \\
    P_\ast &= \sum_{j \geq 1} 31 \cdot 32^{j-1} \phi(7j/2 - u)\,.
\end{align*}

 The explicit coordinate density satisfies $f_d(x)\leq C_\gamma e^{-\gamma x^2/2}$ for each fixed $0<\gamma<1$ and all sufficiently large $d$; integration gives a corresponding bound for $F_d$. These bounds dominate the height sums, including their factors $32^j$, and justify the limits.

\begin{fact}\label{fact:app-upper-numerical}
    We have the numerical bounds $$G < 1.89,\quad H_\ast < 3.05, \quad E_\ast < 0.008, \quad P_\ast < 0.1.$$
\end{fact}
\begin{proof}
    The tails after term $J$ for $H_\ast, E_\ast,P_\ast$ are bounded  (for $a = 7, b = 32, C = ab/(b-1)$) by \begin{gather*}
        H_{> J} \leq b^{-J}\left(\log(1 + C/u) + (J + 1/(b-1)) \log b \right)\\
        E_{> J} \leq \frac{(b-1)b^J e^{-a^2(J+1)^2 / 8}}{2(1 - \rho_E)} \\
        P_{> J} \leq \frac{(b-1)b^J \phi(a(J+1)/2 - u)}{1 - \rho_P} \\
        \rho_E = b e^{-a^2(2J+3)/8}, \quad \rho_P = \rho_E e^{au/2} \,.
    \end{gather*}
    The program \texttt{upper\_simple.py}~\cite{RDR} verifies the stated numerical inequalities using $J = 20$ for $H_\ast$ and $J = 4$ for $E_\ast,P_\ast$.
\end{proof}

\begin{proof}[Proof that $\theta_d(5d) = 0$ for $d$ large]
    By Lemma \ref{lem:app-UB-markov}, Markov's inequality bounds the probability there is an open path from $0$ to distance $\ell_k/\sqrt{h_d}$ by $e^{-\zeta d 32^k}$ where $\zeta > 0$ by Fact \ref{fact:app-upper-numerical} since \begin{align*}
        5 - 1.89 - 3.05 - 5\cdot(0.008) - \frac{225}{128}(0.1)^2 > 0
    \end{align*}
    and so we may take $J$ large, then $\eps$ small, then $d$ large to assure that $\zeta > 0$. Taking $k \to \infty$ shows $\theta_d(5d) = 0$.
\end{proof}

\subsection{Lower bound}
The proof is via the second-moment argument although with spherical increments and a refined version of the various overlap bounds.  

Set $h = 2\pi e, m = d-1$ and $r_+ = \sqrt{r^2 + 1/d}$.  Consider a random walk $\gamma$ in $\R^d$ with independent increments given by $(4,U)/\sqrt{h}$ where $U$ is uniform on the radius $6$ sphere in $\R^m$.  Let $\mathcal{U}_k$ be the radius-$r_+$ tube around the first $k$ edges and define the observable $$Z_k = \E_\gamma e^{\Phi \cdot \Vol_d(\mathcal{U}_k)} \mathbf{1}_{\bX \cap \mathcal{U}_k = \emptyset}\,.$$

For a pair $a,b \in \R^d$ write $[a,b]$ for the line segment between the two points, and  define\begin{align*}
    \mathcal{T}(a,b) &= \{y : \dist(y,[a,b]) \leq r_+\} \\
    \mathcal{F}(a,b) &= \mathcal{T}(a,b) \setminus B(a,r_+) \\ 
    \mathcal{B}(a,b) &= \mathcal{T}(a,b) \setminus B(b,r_+) 
\end{align*}
for the pieces of the tube.  We will need some notation to handle bounds on $Z_k$.  For $U,V$ with $\|U\|_2 = \|V\|_2 = 6$, $j \in \Z$, and $M \in \R^m$, define $$u = (4,U)/\sqrt{h}, \quad v = (4,V)/\sqrt{h}, \quad \delta = (4j,-M)/\sqrt{h}.$$
For two segments, $[0,u]$ and $[\delta,\delta+v]$, write $\mathcal{F}_1 = \mathcal{F}(0,u),\mathcal{B}_1 = \mathcal{B}(0,u), \mathcal{F}_2 = \mathcal{F}(\delta,\delta + v)$, and $\mathcal{B}_2 = \mathcal{B}(\delta,\delta+v)$.  We then define \begin{equation*}
    K_{j,d}(M,U,V) = \frac{1}{2}\left(\Vol_d(\mathcal{F}_1 \cap \mathcal{B}_2) + \Vol_d(\mathcal{B}_1 \cap \mathcal{F}_2) \right)\,.
\end{equation*}

For a path generated with increments $\{U_a\}$, define $A_n = \sum_{a = 0}^{n-1} U_a$ and $S_n = (4n,A_n)/\sqrt{h}$, and note that $\mathcal{U}_k = \bigcup_{0 \leq n < k} \mathcal{T}(S_n,S_{n+1})$.  For another set of increments $\{U_a'\}$ define $A_n', S_n'$ and $\mathcal{U}_k'$ analogously.  Define $M_{n,\ell} = A_n - A_\ell'$.  We now refine the overlap estimate of Lemma \ref{lem:two-replica-by-random-walk}, retaining the moment identity in Eq.~\eqref{eq:second-moment}.

\begin{lemma}\label{lem:app-Zk-moments}
    For every $k \geq 1$, we have $$\E Z_k = 1, \qquad \E Z_k^2 = \E_{\gamma,\gamma'} e^{\Phi \Vol_d(\mathcal{U}_k \cap \mathcal{U}_k')}\,.$$
    In the notation above we have \begin{equation*}
      \Vol_d(\mathcal{U}_k \cap \mathcal{U}_k') \leq 2 \kappa_d r_+^d + \sum_{0 \leq n,\ell < k} K_{\ell - n,d}(M_{n,\ell},U_n,U_\ell')\,.  
    \end{equation*}
\end{lemma}
\begin{proof}
    The proof of the first and second moment identities is identical to the proof of Eqs.~\eqref{eq:Z_k-def} and \eqref{eq:second-moment}.  For the bound on the overlapped volume.  Writing $\mathcal{F}_n = \mathcal{F}(S_n,S_{n+1})$ and $\mathcal{B}_n = \mathcal{B}(S_n,S_{n+1})$, we see that \begin{equation*}
        \mathcal{U}_k = B(S_0,r_+) \cup \bigcup_{n < k} \mathcal{F}_n = B(S_k,r_+) \cup \bigcup_{n < k} \mathcal{B}_n\,.
    \end{equation*}
    Writing $\mathcal{F}_n'$ and $\mathcal{B}_n'$ for analogous primed increments, we see \begin{align*}
        \Vol_d(\mathcal{U}_k \cap \mathcal{U}_k') \leq 2 \kappa_d r_+^d + \sum_{n,\ell < k} \Vol_d(\mathcal{F}_n \cap \mathcal{B}_\ell')\,.
    \end{align*}
    Swapping $U_a$ with $U_a'$ yields the same bound with the term $\Vol_d(\mathcal{B}_n \cap \mathcal{F}_\ell')$ in the summand instead.  Averaging these two and translating each edge pair by $-S_n$ completes the proof.
\end{proof}

We call $j = \ell - n$ the \emph{lag} of the pair of edges with indices $n$ and $\ell$.  The relevant quantity for bounding $K_{j,d}$ is $M_{n,\ell}$.  With this in mind, for fixed lag $j$ set $n_j = \max\{0,-j\}$ and note that the relevant quantities $M_{\cdot,\cdot}$ of lag $j$ are $M_q^{(j)} = A_{n_j + q} - A_{n_j + j + q}'$ for $q \geq 0$.  Note that for fixed $j$ we have the evolution $$M_{q+1}^{(j)} = M_q^{(j)} + U_{n_j + q} - U_{n_j + j + q}'$$
with initial conditions $M_0^{(j)} = -A_j'$ for $j \geq 0$ and $M_0^{(j)} = A_{-j}$ for $j < 0$.  These initial states are independent of the future paired increments. Interchanging the paths gives $K_{-j,d}(M,U,U')=K_{j,d}(-M,U',U)$, so negative-lag bounds follow from positive-lag bounds.  Suppressing the dependence on $j$, we see that this process evolves as $M_{q +1} = M_q + U_q - U_q'$ for i.i.d.\ increments $U_q,U_q'$.  

We will seek to understand the evolution of this process, and so for a transition from a state $M \neq 0$ write $M' = M + U - U'$.  For a non-zero vector $v$, write $\widehat{v} = v/\|v\|_2$.  We now define $G = \mathrm{Gram}(\widehat{M},\widehat{U},\widehat{U'})$ for the $3 \times 3$ Gram matrix generated by these vectors.  A Gram configuration is nonsingular if $\det G > 0$, in which case we define \begin{align*}
    \Delta = \det G, \quad D = \frac{\|M'\|_2^2}{\|M\|^2} \\
    \mathcal{R}(M,U,U') = - \frac{1}{2} \log \Delta + \frac{1}{4} \log D.
\end{align*}

We define the spherical averaging operator $P$ via $(Pf)(M) = \E_U[f(M+U)]$, where $U$ is uniform on the radius $6$ sphere.  Because $-U'$ has the same distribution as $U'$, we note that $\E[f(M')\,|\,M] = (P^2 f)(M)$ for all functions $f$.  Assume $m  > 4$.  Set $p = (m-4)/2$ and $w(M) = \|M\|_2^{-p}$, where we interpret $w(0) = +\infty.$  
\begin{lemma}\label{lem:app_cR-LB}
For $M \neq 0$ we have 
$$\frac{P^2 w(M) }{w(M)} \leq \frac{P w(M)}{w(M)} \leq \left(1 + \frac{36}{\|M\|_2^2}\right)^{-p/4}\,.$$
    Further, for every nonsingular Gram configuration we have \begin{equation*}
    \mathcal{R}(M,U,U') \geq \frac{1}{8}\log(1 + 36/ \|M\|_2^2)\,.
    \end{equation*}
\end{lemma}
\begin{proof}
    For $a > 0$, we introduce a regularization $w_a = (a + \|M\|_2^2)^{-p/2}$, and set $F(s)$ to be the spherical mean at radius $s$ of $w_a$ about $M$.  By direct computation we have $- \Delta w_a \geq p(m - 2 - p)w_a / (a + \|M\|_2^2)$ and so Jensen gives the lower bound \begin{equation*}
        -\frac{F'(s)}{F(s)} \geq \frac{p(m - 2 - p)s}{m(a + \|M\|_2^2 + s^2)}\,.
    \end{equation*}
    Integrating to $s = 6$ and taking $a \downarrow 0$ yields the stated upper bound.  Superharmonicity gives the bound $P^2 w \leq Pw.$ 

    For the lower bound on $\mathcal{R}$, write $q = \|M\|_2, q' = \|M'\|_2, c = \langle \widehat{U},\widehat{U}'\rangle$, and $Z = U - U'$.  Let $M_{||}$ denote the projection of $M$ onto $\mathrm{span}(\{U,U'\})$, and decompose $M = M_{||} + M_\perp$ to be the orthogonal decomposition; set $z = \|M_\perp\|_2$.  Nonsingularity of the the Gram configuration implies $z > 0$ and $|c| < 1$.  Factoring the Gram determinant gives $\Delta = (1 - c^2)z^2 / q^2$, and so \begin{equation*}
        e^{4 \mathcal{R}} = \frac{q^2 (q')^2}{(1 - c^2)^2 z^4}\,.
    \end{equation*}
    We then lower bound \begin{align*}
        q^2(q')^2 &= (z^2 + \|M_{||}\|_2^2)(z^2 + \|M_{||} + Z\|_2^2) \\
        &\geq z^4 + \frac{1}{2}z^2\|Z\|_2^2\,.
    \end{align*}
    Using $z \leq q, \|Z\|^2 = 72(1 - c)$ and $|c| < 1$, we then obtain \begin{align*}
        e^{4 \mathcal{R}} \geq \frac{1 + 36(1 - c)/z^2}{(1 - c^2)^2} \geq 1 + 18/q^2\,.
    \end{align*}
    Using the elementary inequality $(1 + 18/q^2)^2 \geq 1 + 36/q^2$ and taking logarithms complete the proof.
\end{proof}

Our goal now is to handle the case of small $j$.  We will show $K_{j,d}$ converges to a limit that we will bound in terms of an explicit integral.  Similarly, we will lower bound $\mathcal{R}.$  We first introduce quite a bit of notation.  For nonsingular Gram data $M,U,U'$, write $e_\pm = (U \pm U')/ \|U \pm U'\|_2$, and put $C=M+(U-U')/2$, $x=\langle C,e_-\rangle$, $y=\langle C,e_+\rangle$, $C_\perp=C-xe_--ye_+$, $z=\|C_\perp\|_2>0$, and $e_\perp=C_\perp/z$. Thus \begin{equation*}
    M + (U -U')/2 = x e_- + y e_+ + z e_\perp 
\end{equation*}
with $z  > 0$ by nonsingularity.  Set \begin{gather*}
    s^2 = y^2 + z^2, \quad w = y/s, \quad \tau = \langle \widehat{U},\widehat{U}'\rangle \\
    k^2 = 18(1 + \tau),\quad \beta^2 = 18(1 - \tau), \quad \alpha^2 = 16 + k^2. \\
    A = (16j - kws)/\alpha + \alpha(u - t), \quad B = \beta(t + u - 1) \\
    Z^2 = s^2(1 - w^2) + 16(jk + sw)^2 / \alpha^2 \\
    T = A^2 + B^2 + Z^2 \\
    \mathcal{A}_j = \int_0^1\int_0^1 \frac{e^{-T/8}}{\sqrt{2\pi  T}} \left(\frac{\beta^2(A^2 + Z^2)}{2T} + \frac{\alpha^2 A^2}{8} \right)\,dt\,du\,.
\end{gather*}
We also define\begin{align*}
    \varphi(q) &= \begin{cases}
        \frac{1}{2}\log(1 + q), & q \leq 1 \\
        \frac{1}{4}\log(4q), \quad q \geq 1
    \end{cases} \\
    \mathcal{R}_\ast &= -\frac{1}{2} \log\left((1 - \tau^2)(1 - w^2) \right) + \varphi(18(1 - \tau)/s^2)\,.
\end{align*}

\begin{lemma}\label{lem:app-Kj-limit-bound}
    For each fixed $j$, the kernels $K_{j,d}$ converge uniformly on compact nonsingular Gram families to a limit $K_j$.  This limit along with the quantity $\mathcal{R}$ satisfy \begin{equation*}
        K_j \leq \mathcal{A}_j(s,\tau,w), \quad \mathcal{R} \geq \mathcal{R}_\ast(s,\tau,w)\,.
    \end{equation*}
\end{lemma}
\begin{proof}
    Write $e = (4,U)$, $f = (4,U')$, and $Y = (4j,-M) + uf - te$ for variables $t,u \in [0,1]$, which indicate positions along the edges $e$ and $f$ respectively.  Write $p_t = te / \sqrt{h}$ and $q_u = ((4j,-M) + uf)/\sqrt{h}$ for these points on the edges.  A point in a forward part $\mathcal{F}_1$ corresponding to the edge $e$  has a unique entry time $t$ into a ball of radius $R = r_+$ centered at $p_t$; similarly, a point in the backward part $\mathcal{B}_2$ corresponding to $f$ has a unique exit time $u$.  For a point on the boundaries of the $R$-balls centered at $p_t$ and $q_u$ simultaneously, let $n_1$ and $n_2$ be the corresponding outward normal vectors.  Set $\nu_1 = \langle e/\sqrt{h}, n_1\rangle$ and $\nu_2 = \langle f/\sqrt{h},n_2\rangle.$  The crossing-speed factor $(\nu_1)_+(-\nu_2)_+$ is a coarea weight for $\mathcal F_1\cap\mathcal B_2$: its integration against the common-boundary measure and the two crossing times gives the volume. Averaging the two entry-exit weights gives $$\chi(\nu_1,\nu_2) := \frac{|\nu_1 \nu_2| - \nu_1\nu_2}{4} \leq (\nu_1 - \nu_2)^2 / 8\,.$$
    Write $\ell = \|Y\|_2 / \sqrt{h}$ and $\rho^2 = R^2 - \ell^2/4.$  Letting $\omega$ be uniform on the unit sphere in $Y^\perp$ we then have the identity \begin{align*}
        K_{j,d} &= \int_0^1 \int_0^1 J_d(Y) \E_\omega \chi(\nu_1,\nu_2)\,dt\,du\\
        J_d(Y) &= (d - 1) \kappa_{d-1} R^2 \rho^{d - 3}/\ell\,,
    \end{align*}
    where the common boundary point is $te/\sqrt{h} + Y/(2\sqrt{h}) + \rho\omega.$  Note that the integrand is zero for $\ell \geq 2R$.  The normals at this point may then simplify $\nu_1 - \nu_2$ as \begin{equation*}
        \nu_1 - \nu_2 = \frac{1}{R}\left(\frac{\langle e + f, Y \rangle}{2h} + \frac{\rho}{\sqrt{h}} \langle e - f,\omega \rangle \right)\,.
    \end{equation*}
    Note that $\E \omega = 0$ and $\E \langle v,\omega\rangle^2 = \|v_{Y^\perp}\|_2^2 / (d - 1)$ for all $v$.  We will use the upper bound on $\chi$ along with these bounds but first note that Stirling's formula gives \begin{gather*}
        h R^2 / d \to 1, \quad \kappa_{d-1} R^{d - 1} / \sqrt{h} \to 1/\sqrt{2\pi} \\
        (\rho/R)^{d-1} \to e^{-\|Y\|_2^2 / 8}\,.
    \end{gather*}
    Taking the limit yields convergence of $K_{j,d} \to K_j$ (where uniform fourth-moment bounds along with a uniform lower bound on $\|Y\|_2$ yield uniform convergence on compact sets of nonsingular families). Using the upper bound on $\chi(a,b) \leq (a - b)^2 / 8$ yields \begin{align*}
        K_j &\leq \frac{1}{\sqrt{2\pi}} \int_0^1 \int_0^1 \frac{e^{-\|Y\|_2^2 / 8}}{\|Y\|_2} \\
        &\quad \times \left(\frac{\|(e - f)_{Y^\perp}\|_2^2}{8} + \frac{\langle e + f,Y\rangle^2}{32} \right)\,dt\,du.
    \end{align*}

    We now need to change variables to massage this bound into the claimed upper bound on $K_j$.  
    Set $v = u -t$ and $r = t + u - 1$, and note that $$Y = (4j + 4v, (kv - y)e_+ - (x + \beta r) e_- - z e_\perp)\,.$$
    Projection onto the orthogonal unit vectors $a_0 = (4,k e_+)/\alpha$ and $(0,e_-)$ gives $\langle a_0,Y \rangle = A$, the second coordinate $-(x + B)$ and the remaining length squared is $Z^2$.  This gives the parameterizations \begin{align*}
        T_x &:= \|Y\|_2^2 = A^2 + (x + B)^2 + Z^2 \\
        \|(e -f)_{Y^\perp}\|_2^2 &= 4\beta^2 (A^2 + Z^2)/T_x \\
        \langle e + f,Y\rangle^2 &= 4\alpha^2 A^2\,.
    \end{align*}
    Substitution into the upper bound on $K_j$ yields $\mathcal{A}_j$ with $B$ replaced by $x + B$.  For fixed $v=u-t$, the remaining integrand is an even function $F_v(x+B)$, nonincreasing in $|x+B|$, and $B$ ranges symmetrically over $[-L,L]$, where $L=\beta(1-|v|)$. Thus, for $x\geq0$, $\frac{d}{dx}\int_{-L}^{L}F_v(x+b)\,db=F_v(x+L)-F_v(x-L)\leq0$. The integral is even in $x$, so its maximum occurs at $x=0$, proving $K_j\leq\mathcal A_j(s,\tau,w)$.

    For the lower bound on the rate $\mathcal{R}$, we note that the same coordinate system gives \begin{gather*}
        \|M\|_2^2 = s^2 + (x - \beta)^2, \quad \|M'\|_2^2 = s^2 + (x + \beta)^2 \\
        \Delta = (1 - \tau^2)z^2 /\|M\|_2^2\,.
    \end{gather*}
    Setting $P(x) = (s^2 + (x - \beta)^2)\cdot (s^2 + (x + \beta)^2)$ we may write \begin{equation*}
        \mathcal{R} = -\frac{1}{2} \log\left[(1 - \tau^2)(1 -w^2) \right] + \frac{1}{4} \log(P(x) / s^4)\,.
    \end{equation*}
    Expanding $P$ as a quadratic in $x^2$ and noting that its minimum occurs at $x^2 = \max\{0,\beta^2 - s^2\}$ completes the proof.
\end{proof}

We now prove bounds for $j = 0,1,2$: \begin{lemma}\label{lem:app-global-UB-K_j-small}
    We have the uniform bounds $K_j < C_j$ for all $j \in \{0,1,2\}$, where $C_0 = 3, C_1 = 6/5$, and $C_2 = 1/20$.
\end{lemma}
\begin{proof}
    For $j = 0$, $K_0$ is bounded by the limiting volume of adding one edge the volume is bound by $\sqrt{52/(2\pi)} < 3.$  For $j \in \{1,2\}$, split the intersection at the longitudinal midplane between its nearer endpoints and extend the two edges away from that plane to rays.  Set $a = 2/\sqrt{13}, \sigma = 3/\sqrt{13},z = 2(j-1)$.  The two ray caps have total limiting volume \begin{equation*}
        2Q(z) + \frac{2}{a} \int_z^\infty (q - z)\phi(q)[\sigma \phi(aq/\sigma) - aqQ(aq/\sigma)]\,dq.
    \end{equation*}
    For $z = 0$, we may use polar coordinates to obtain $1 + ((3/2) - \arctan(3/2))/\pi < 6/5$.  For $z = 2$, we may omit the negative term and integrate to obtain $$2Q(2) + \frac{\sigma^3}{\pi a} e^{-2/\sigma^2} - \frac{4\sigma^2}{a\sqrt{2\pi}} Q(2/\sigma) < 1/20.$$
    The claimed numerical inequalities are checked in program \texttt{global\_caps.cpp} \cite{RDR}.
\end{proof}

We require some basic numerical bounds once again.  \begin{fact}\label{fact:app-numerical-2}
    Set $A_0 = 6/5$, $A_1 = 3/8$, and $A_2 = 1/64$.  For $j \in \{0,1,2\}$ we have $$\min\{C_j,\mathcal{A}_j(s,\tau,w)\} \leq A_j \mathcal{R}_\ast(s,\tau,w)$$
    on the set $1/200 \leq s \leq 24$, $|\tau| , |w| \leq 1 $, and $(1 - \tau^2)(1 - w^2) \geq 2^{-10}.$
\end{fact}
\begin{proof}
    The program \texttt{verify\_lower.py}~\cite{RDR} verifies this inequality by exhaustive interval subdivision for the integral defining $\mathcal{A}_j$.  Let $f(t,u)$ denote the scalar integrand defining $\mathcal A_j$ (distinct from the edge vector $f$ above), and let $Q\subset[0,1]^2$ be a time rectangle of area $\Vol(Q)$ before the change of variables. On a rectangle with sides $h_1,h_2$, the midpoint-rule error per area is at most $(h_1^2 \sup|f_{tt}| + h_2^2 \sup |f_{uu}|)/24$.  Interval differentiation and interval bounds handle both terms.  For rectangles near $T = 0$, the change to coordinates $(A,B)$ has Jacobian $2\alpha \beta$ and $\int_Q T^{-1/2} \leq 2 \sqrt{\pi \Vol(Q)/(2\alpha \beta})$, which provides an integrable bound.  The rate is minimized over a parameter box by its largest $s$, smallest $|w|,$ and $\tau$ value \begin{equation*}
        \tau_\ast(s) = \begin{cases}
            1/3 & s^2 \leq 12 \\
            \frac{18}{s^2 + 18 + \sqrt{(s^2 + 18)^2 - 324}} & s^2 \geq 12
        \end{cases}
    \end{equation*}
        restricted to the allowed $\tau$ interval.  For $j=0$, swapping $t,u$ gives $\mathcal A_0(s,\tau,-w)=\mathcal A_0(s,\tau,w)$, and $\mathcal R_\ast$ is even in $w$. Hence the $168$ initial boxes with $w\geq0$ for $j=0$, together with $504$ boxes each for $j=1,2$ covering both signs, cover the full parameter domain. The stored computation covers these $1,176$ initial boxes and records no unresolved boxes.
\end{proof}

We now upper bound $K_j$ in terms of $\mathcal{R}$ using this numerical certificate.
\begin{lemma}\label{lem:app-numerical-UB-small-j}
    For $j \in \{0,1,2\}$ and every nonsingular Gram configuration, we have $K_j \leq A_j \mathcal{R}$, where $A_j$ is as in Fact \ref{fact:app-numerical-2}.
\end{lemma}
\begin{proof}
    Combining Lemmas \ref{lem:app-Kj-limit-bound} and \ref{lem:app-global-UB-K_j-small} with Fact \ref{fact:app-numerical-2} handles the compact parameter domain defined in Fact \ref{fact:app-numerical-2}.  For $s \leq 1/200$, note that $\mathcal{R}_\ast > 7/2$; if $(1 - \tau^2)(1 - w^2) \leq 2^{-10}$ we have $\mathcal{R}_\ast \geq 5 \log 2 > 10/3$, thus handling these two cases.

    For $s \geq 24$, write $q = \|M\|_2 \geq 24$ and cover each segment at its endpoints and midpoints by balls of squared radius $r_+^2 + 52/(16h)$.  Bounding each lens in its midpoint ball gives the bound $K_j \leq 9 e^{13/8} e^{-(q-12)^2/8}.$  Using Lemma \ref{lem:app_cR-LB} provides \begin{equation*}
        \frac{K_j}{\mathcal{R}} \leq 2 e^{\frac{13}{8}}(q^2 + 36) e^{-(q -12)^2/8} \leq 1224 e^{-\frac{131}{8}} < \frac{1}{64}\,,
    \end{equation*}
    where we used that the middle expression decreases for $q \geq 24$.  The scalar comparisons are checked in the program \texttt{exterior\_bounds.py}~\cite{RDR}.
\end{proof}

\begin{lemma}\label{lem:app-large-j}
    For all large $d$, uniformly in $j \geq 3$ and in the $M,U,U'$ we have $$K_{j,d}(M,U,U') \leq \frac{13000^{3-j}}{2800} \cdot e^{-\|M\|_2^2 / 162}\,.$$
\end{lemma}
\begin{proof}
    With $Y = (4j,-M) + u(4,U') - t(4,U)$, as in Lemma \ref{lem:app-Kj-limit-bound}, using the squared edge length $52/h$ gives \begin{gather*}
        K_{j,d} \leq 21 \int_0^1 \int_0^1 G_b(\|Y\|_2) \,dt\,du \\
        G_b(z) = e^{-b z^2 / 8}\left(\frac{1}{2z} + \frac{z}{8} \right)\,,
    \end{gather*}
    where $b = (d -3)/(h r_+^2) \to 1$, and the prefactor tends to $52/\sqrt{2\pi} \leq 21$.  This bound is uniform in $M,U,U'$ and $j$.  For $b \geq 1/2$, the quantity $e^{b z^2 / 16} G_b(z)$ decreases, which can be checked by computing the derivative.  Writing $q = \|M\|_2$ as before, the squared transverse separation is at least $(q - 6(t+u))_+^2 \geq q^2/10 - 4(t+u)^2.$ For $j \geq 3$, we then have \begin{gather*}
        K_{j,d} \leq 21 e^b I_j(b) e^{-bq^2 / 160} \\
        I_j(b) = \int_0^\infty w G_b(4(j-1+w))\,dw\,.
    \end{gather*}
    Here $w=1+u-t$ has density $\min(w,2-w)$ on $[0,2]$, bounded by $w$; extending the nonnegative integral to $[0,\infty)$ gives $I_j$. 
    Since $G_b(z + 4)/G_b(z) \leq (3/2)e^{-10b}$ for $z \geq 8$, we have $I_j \leq I_3 [(3/2) e^{-10b} ]^{j-3}$.  Integration by parts along with Mills' inequality $Q(t) \leq \phi(t)/t$ and $\int_x^\infty e^{-z} z^{-1}\,dz \geq e^{-x}/(x+1)$ shows $$I_3(b) \leq e^{-8b} \left(\frac{b+1}{64\cdot b^2} - \frac{1}{8(8b+1)} \right)\,.$$
    Defining $b_\pm = 1 \pm \frac{1}{200}$ we see \begin{gather*}
        21 e^{b_+}I_3(b_-) < 1/2800 \\
        \frac{3}{2} e^{-10 b_-} < 1/13000, \qquad b_-/160 > 1/162\,.
    \end{gather*}
    These scalar checks---verified in the program \texttt{tail\_simple.py}~\cite{RDR}---complete the proof.
\end{proof}

We are now ready to show that we can handle bounds for small $j$.
\begin{lemma}\label{lem:app-exponential-moment-small-j}
    For each $j \in \{0,1,2\}$, fix any $a_j > A_j$.  Then for all sufficiently large $d$ we have \begin{equation*}
        \sup_M \E_M \exp\left(\frac{m}{a_j} \sum_{n \geq 0} K_{j,d}(M_n,U_n,U_n') \right) < \infty\,.
    \end{equation*}
\end{lemma}
\begin{proof}
    Fix $j \in \{0,1,2\}$ throughout. 
    Put $\mu = m/a_j$, $q = \|M\|_2$, and as before set $w(M) = q^{-p}$.  Define the operator $$T_j f(M) = \E_{U,U'}[ e^{\mu K_{j,d}(M,U,U')} f(M + U - U')]\,.$$
    Setting $a = \langle \widehat{M}, \widehat{U}\rangle$, $b = \langle \widehat{M},\widehat{U}'\rangle$, and $c = \langle \widehat{U},\widehat{U}'\rangle$, we see that this trio have joint density $c_m \Delta^{(m-4)/2}$, where $c_m = O(m^{3/2})$.  Since $w(M')/w(M) = D^{-(m-4)/4}$, we have \begin{equation*}
        \frac{T_j w(M)}{w(M)} = c_m \int e^{-(m-4) \mathcal{R} + \mu K_{j,d}} \,da\,db\,dc\,,
    \end{equation*}
    where the integral is over positive definite Gram configurations.  Fix $\overline a\in(A_j,a_j)$ and an annulus $\eps\leq q\leq L$, with $\eps$ small and $L$ large, to be chosen below.  Lemma \ref{lem:app_cR-LB} gives $\mathcal{R} \geq r_0 := \log(1 + 36/L^2)/8 > 0$.  Lemmas \ref{lem:app-Kj-limit-bound} and \ref{lem:app-numerical-UB-small-j} give $K_{j,d}\leq\overline a\mathcal R$ for all large $d$ on the whole annulus. Indeed, use compact convergence away from $\Delta=0$; near that boundary use the uniform bound $K_{j,d}\leq C$ from the one-edge volume and $\mathcal R\geq-\frac14\log\Delta$, since $\Delta/D\leq1$ is the normalized Gram determinant with $M$ replaced by $M'$. The angular domain has volume at most $8$, so the integral is bounded by $8c_m e^{-[(m-4)-m\overline a/a_j]r_0}$, and for large $m$ we have \begin{equation*}
        \frac{T_j w(M)}{w(M)} \leq \frac{1}{2}\,.
    \end{equation*}
    To handle the radial ends, the one-edge volume gives $K_{j,d}\leq C$ uniformly for all large $d$. The three-ball cover used in Lemma~\ref{lem:app-numerical-UB-small-j}, applied before taking the dimension limit, also gives $K_{j,d}\leq B(q):=C_1e^{-q^2/64}$ for $q\geq24$: the inflated ball volume is uniformly bounded, the center separation is at least $(q-12)/\sqrt h$, and the lens bound has exponent at most $-(q-12)^2/16\leq-q^2/64$. Lemma~\ref{lem:app_cR-LB} therefore gives
    \begin{equation*}
        T_jw/w\leq e^{\mu B(q)}(1+36/q^2)^{-(m-4)/8},
    \end{equation*}
    using $B(q)=C$ near zero. Choose $\eps$ small enough that this is at most $1/2$ for $q\leq\eps$. Choose $L$ large enough that $\mu B(q)\leq(m-4)\log(1+36/q^2)/16$ for $q\geq L$. These choices work uniformly for large $m$ and give $T_jw/w\leq(1+36/q^2)^{-(m-4)/16}<1$ at infinity.

    Let $g = T_j 1 - 1$ and $h_0 = w - T_j w > 0$.  For $q\leq L$, the preceding bounds give $h_0\geq w/2$ and $g\leq e^{\mu C}-1$. At infinity, they give $h_0\geq c(d)q^{-p-2}$ whereas $g\leq C(d)e^{-q^2/64}$. Therefore, $g/h_0$ is bounded.  We then proceed in a similar manner to Corollary \ref{cor:exponential-moment-3D}. In particular, setting $A \geq \sup_{M \neq 0} g(M)/h_0(M)$ then we have $T_j(1 + Aw) \leq 1 + Aw$ and iteration first gives $T_j^N1(M)\leq1+Aw(M)$ for $M\neq0$. One additional transition, using $K_{j,d}\leq C$, then gives the uniform bound (including $M=0$) $$T_j^N 1(M) \leq e^{O(\mu)}(1 + A \sup_X P^2 w(X)) < \infty\,, $$
    because the density of $U-U'$ is radial decreasing and behaves like $|z|^{-1}$ near zero. Its convolution with $w$ is maximal at zero and finite because $p<m-1$.  Since $T_j^N 1(M)$ is the exponential moment of the first $N$ contributions, the monotone convergence theorem completes the proof.
\end{proof}

The case of $j \geq 3$ will be handled by the following Lemma. \begin{lemma}\label{lem:app-large-j-uniform-bound-final}
    For $m \geq 400$ we have $$\sup_M \E_M \exp\left(\frac{m}{17} \sum_{n \geq 0} e^{-\|M_n\|_2^2 / 162} \right) < \infty\,.$$
\end{lemma}
\begin{proof}
    We apply Lemma \ref{lem:app_cR-LB} directly to the multiplicative potential $\exp(m e^{-\|M\|_2^2 /162}/17)$.  With $B = 2/9$ we have $\inf_{q > 0} e^{B/q} \log(1 + q) \geq B e^{1-B}$ and $0.99 \cdot B e^{1-B}/ 8 > 1/17$.  This shows that the weighted transition is contracting.  The argument considering $1 + Aw$ in the proof of Lemma \ref{lem:app-exponential-moment-small-j} shows the required bound; the necessary scalar comparison for this lemma is verified in the program \texttt{tail\_simple.py}~\cite{RDR}.
\end{proof}

\begin{proof}[Proof of lower bound]
Set $\Phi = d/2$ and group the contributions by their lag writing \begin{equation*}
    H_{j,k} = \sum_{\substack{0 \leq n < k \\ 0 \leq n + j < k}} K_{j,d} (M_{n,n+j},U_n,U_{n+j}'), \quad j \in \Z\,,
\end{equation*}
and note that the sum only has finitely many non-zero terms.  

Choose $a_i > A_i$ for $i \in \{0,1,2\}$, and define the positive weights \begin{equation*}
    b_j = \begin{cases}
        a_{|j|}, & |j| \leq 2 \\
        \frac{17}{2800} \cdot 13000^{3-|j|}, & |j| \geq 3\,.
    \end{cases}
\end{equation*}
Note that the tail weights sum to a value $< 1/80$.  Taking $a_i$ sufficiently close to $A_i$, we may then guarantee that $$W = \sum_j b_j < 2\,.$$
For each given lag, the difference process starts at a displacement independent of its future paired increments.  For $|j| \leq 2$, we condition on that displacement and apply Lemma \ref{lem:app-exponential-moment-small-j}.  For $|j| \geq 3,$ we may divide the conclusion of Lemma \ref{lem:app-large-j} by $b_j$, thus allowing the use of Lemma \ref{lem:app-large-j-uniform-bound-final}.  Therefore, there is some finite $C_d$ so that for all $j$ and $k$ we have $$\E e^{m H_{j,k}/b_j} \leq C_d\,,$$
where the expectation is over paths.  This now sets us up to bound the second moment.

Define $\rho_j =b_j/W$ so $\sum \rho_j = 1$.  Since $W < 2$ and $m = d-1$ we have $\Phi W < m$ for all $d$ sufficiently large.  Jensen's inequality now gives $$\E e^{\Phi \sum_{j} H_{j,k}} \leq C_d\,.$$
Combining with Lemma \ref{lem:app-Zk-moments}, we see that $\E Z_k = 1$ and $\E Z_k^2 \leq e^{2\Phi \kappa_d r_+^d} C_d < \infty$.  By the Paley--Zygmund inequality together with K\"onig's lemma used to prove Proposition \ref{prop:lower-bound} the proof is complete.
\end{proof}

\bibliography{Bib}

\end{document}